\documentclass{article}
\usepackage{graphicx}
\usepackage{pifont}
\usepackage{geometry}
\usepackage[cmex10]{amsmath}
\usepackage{array}
\usepackage{multirow}
\usepackage{bm}
\usepackage{authblk}
\usepackage{mdwmath}
\usepackage{xcolor,soul,framed}
\usepackage{cite}
\usepackage{eqparbox}
\usepackage{url}
\usepackage{amsmath}
\usepackage{multirow}
\usepackage{amssymb}
\usepackage{textcomp}
\usepackage{bm}
\usepackage{dblfloatfix}
\usepackage{svg}
\usepackage{anyfontsize}

\definecolor{darkgreen}{RGB}{0,102,51}

\newgeometry{hmargin={20mm,20mm}}   %

\begin{document}

\title{Distribution System State and Impedance
Estimation Augmented with Carson’s Equations}

\author[1,2]{Marta Vanin}
\author[3]{Frederik Geth}
\author[4]{Rahmat Heidari}
\author[1,2]{Dirk Van Hertem}

\affil[1]{{Department of Electrical Engineering, KU Leuven, Belgium}}
\affil[2]{{Etch-EnergyVille, Genk, Belgium}}
\affil[3]{{GridQube, Brisbane, Australia}}
\affil[4]{{CSIRO Energy, Newcastle, Australia}}

\date{}

\maketitle

\newcommand{\branches}{$\mathcal{E}$}
\newcommand{\branchesm}{\mathcal{E}}
\newcommand{\branchessp}{$\mathcal{E}^{1ph}$}
\newcommand{\branchesspm}{\mathcal{E}^{1ph}}
\newcommand{\branchestp}{$\mathcal{E}^{3ph}$}
\newcommand{\branchestpm}{\mathcal{E}^{3ph}}
\newcommand{\branchesi}{$\mathcal{E}_i$}
\newcommand{\branchesim}{\mathcal{E}_i}
\newcommand{\buses}{$\mathcal{N}$}
\newcommand{\busesm}{\mathcal{N}}
\newcommand{\conds}{$\phi$}
\newcommand{\condsm}{\phi}
\newcommand{\condsij}{$\phi_{ij}$}
\newcommand{\condsijm}{\phi_{ij}}
\newcommand{\gens}{$\mathcal{G}$}
\newcommand{\gensm}{\mathcal{G}}
\newcommand{\gensi}{$\mathcal{G}_i$}
\newcommand{\gensim}{\mathcal{G}_i}
\newcommand{\linecodessp}{$\mathcal{LC}^{1ph}$}
\newcommand{\linecodesspm}{\mathcal{LC}^{1ph}}
\newcommand{\linecodestp}{$\mathcal{LC}^{3ph}$}
\newcommand{\linecodestpm}{\mathcal{LC}^{3ph}}
\newcommand{\loads}{$\mathcal{L}$}
\newcommand{\loadsm}{\mathcal{L}}
\newcommand{\loadsi}{$\mathcal{L}_i$}
\newcommand{\loadsim}{\mathcal{L}_i}
\newcommand{\meas}{$\mathcal{M}$}
\newcommand{\measm}{\mathcal{M}}
\newcommand{\pseudom}{$\mathcal{P}$}
\newcommand{\mpseudom}{\mathcal{P}}
\newcommand{\rbuses}{$\mathcal{R}$}
\newcommand{\rbusesm}{\mathcal{R}}
\newcommand{\shunts}{$\mathcal{S}$}
\newcommand{\shuntsm}{\mathcal{S}}
\newcommand{\shuntsi}{$\mathcal{S}_i$}
\newcommand{\shuntsim}{\mathcal{S}_i}
\newcommand{\timeseries}{$\mathcal{T}$}
\newcommand{\timeseriesm}{\mathcal{T}}


\newcommand{\z}{$\mathbf{z}$}
\newcommand{\h}{$\mathbf{h}$}
\newcommand{\hhm}{$\mathbf{h}_m$}
\newcommand{\errv}{$\boldsymbol{\eta}$}
\newcommand{\va}{$\angle U$}
\newcommand{\vm}{$|U|$}
\newcommand{\vmjp}{$|U_{j,p}|$}
\newcommand{\vajp}{$\angle U_{j,p}$}
\newcommand{\vi}{$U^{\text{im}}$}
\newcommand{\vr}{$U^{\text{re}}$}
\newcommand{\ca}{$\angle I $}
\newcommand{\cax}{$\angle I_c $}
\newcommand{\cix}{$I^{\text{im}}_c$}
\newcommand{\crx}{$I^{\text{re}}_c$}
\newcommand{\cmx}{$|I_c|$}
\newcommand{\px}{$P_c$}
\newcommand{\qx}{$Q_c$}
\newcommand{\w}{$W$}

\newcommand{\cixm}{I^{\text{im}}_c}
\newcommand{\crxm}{I^{\text{re}}_c}
\newcommand{\cim}{I^{\text{im}}}
\newcommand{\crm}{I^{\text{re}}}
\newcommand{\crmlijt}{\textbf{I}^{\text{re}}_{\text{lij,t}}}
\newcommand{\crmit}{\textbf{I}^{\text{re}}_{\text{i,t}}}
\newcommand{\crmut}{\textbf{I}^{\text{re}}_{\text{u,t}}}
\newcommand{\cimlijt}{\textbf{I}^{\text{im}}_{\text{lij,t}}}
\newcommand{\cimit}{\textbf{I}^{\text{im}}_{\text{i,t}}}
\newcommand{\cimut}{\textbf{I}^{\text{im}}_{\text{u,t}}}
\newcommand{\urjt}{\textbf{U}^{\text{re}}_{\text{j,t}}}
\newcommand{\urit}{\textbf{U}^{\text{re}}_{\text{i,t}}}
\newcommand{\uijt}{\textbf{U}^{\text{im}}_{\text{j,t}}}
\newcommand{\uiit}{\textbf{U}^{\text{im}}_{\text{i,t}}}
\newcommand{\cmxm}{|I|_c}
\newcommand{\vim}{U^{\text{im}}}
\newcommand{\vrm}{U^{\text{re}}}
\newcommand{\cam}{\angle I}
\newcommand{\caxm}{\angle I_c}
\newcommand{\vvm}{U}
\newcommand{\vmm}{|U|}
\newcommand{\vmmpfi}{|U_i^{\text{pf}}|}
\newcommand{\vmmsei}{|U_i^{\text{se}}|}
\newcommand{\vam}{\angle U}
\newcommand{\pxm}{P_c}
\newcommand{\qxm}{Q_c}
\newcommand{\p}{P}
\newcommand{\qm}{Q}
\newcommand{\wm}{W}

\newcommand{\idcondone}{p}
\newcommand{\idcondtwo}{q}
\newcommand{\idline}{l}
\newcommand{\setCond}{\mathcal{P}}

\newcommand{\imagunit}
{\textcolor{black}{j}}

\newcommand{\Ziibar}
{\textcolor{black}{Z^{\text{nom}}_{\idcondone\idcondone, k}}}
\newcommand{\Riibar}
{\textcolor{black}{R^{\text{nom}}_{\idcondone\idcondone,k}}}
\newcommand{\Xiibar}
{\textcolor{black}{X^{\text{nom}}_{\idcondone\idcondone,k}}}
\newcommand{\Zijbar}
{\textcolor{black}{Z^{\text{nom}}_{\idcondone\idcondtwo, k}}}
\newcommand{\Rijbar}{\textcolor{black}{R^{\text{nom}}_{\idcondone\idcondtwo,k}}}
\newcommand{\Xijbar}{\textcolor{black}{X^{\text{nom}}_{\idcondone\idcondtwo,k}}}

\newcommand{\Zii}{\textcolor{black}{Z_{l,\idcondone\idcondone}}}
\newcommand{\Rii}{\textcolor{black}{R_{l,\idcondone\idcondone}}}
\newcommand{\Xii}{\textcolor{black}{X_{l,\idcondone\idcondone}}}
\newcommand{\Zij}{\textcolor{blue}{Z_{l,\idcondone\idcondtwo}}}
\newcommand{\Rij}{\textcolor{black}{R_{l,\idcondone\idcondtwo}}}
\newcommand{\Xij}{\textcolor{black}{X_{l,\idcondone\idcondtwo}}}

\newcommand{\relperm}
{\textcolor{black}{\mu^{\text{rel}}}}

\newcommand{\ri}{\textcolor{black}{r_{\idcondone, k}}}

\newcommand{\gmri}{\textcolor{black}{f^{\text{GMR}}_{\idcondone, k}}}
\newcommand{\dij}{\textcolor{black}{D_{\idcondone\idcondtwo,k}}}

\newcommand{\xvec}{\textcolor{black}{\mathbf{x}}}
\newcommand{\yvec}{\textcolor{black}{\mathbf{y}}}
\newcommand{\xxi}{\textcolor{black}{x_{\idcondone}}}
\newcommand{\xj}{\textcolor{black}{x_{\idcondtwo}}}
\newcommand{\yi}{\textcolor{black}{y_{\idcondone}}}
\newcommand{\yj}{\textcolor{black}{y_{\idcondtwo}}}

\newcommand{\coefone}
{\textcolor{black}{c_1}}
\newcommand{\coeftwo}
{\textcolor{black}{c_2}}

\newcommand{\rhoi}{\textcolor{black}{\rho_{\idcondone, k}}}
\newcommand{\Ai}{\textcolor{black}{A_{\idcondone, k}}}
\newcommand{\lenl}{\textcolor{black}{l_{\idline}}}

\newcommand{\Aimin}{\textcolor{red}{A^{\text{min}}_{\idcondone}}}
\newcommand{\Aimax}{\textcolor{red}{A^{\text{max}}_{\idcondone}}}

\begin{abstract}
The impedances of cables and lines used in (multi-conductor) distribution networks are usually unknown or approximated, and may lead to problematic results for \textit{any} physics-based power system calculation, e.g., (optimal) power flow. Learning parameters from time series data is one of the few available options to obtain improved impedance models. This paper presents an approach that combines statistical learning concepts with the exploitation of domain knowledge, in the form of Carson's equations, through nonlinear mathematical optimization. The proposed approach derives impedance matrices for up-to-four-wire systems, using measurement data like those obtained from smart meters. Despite the lack of phasor measurements, the low signal-to-noise ratio of smart meter measurements, and the inherent existence of multiple equivalent solutions, our method produces good quality impedance models that are fit for power system calculations, significantly improving on our previous work both in terms of accuracy and computational time.
\end{abstract}

\textbf{keywords -}
Carson's equations,  distribution system state estimation, impedance estimation, parameter estimation.

\section{Introduction}\label{s:introduction}

Carson's equations~\cite{Carson1926} are considered the best impedance calculation practice; they are featured in distribution network analysis textbooks and manuals, e.g.,~\cite{kersting2018distribution}, and are natively supported in popular distribution system tools, e.g., OpenDSS 
and CYME. 
However, it is usually challenging to use these equations to calculate impedances for existing distribution networks (DNs): \textit{1)} information on the precise cable/line types in use at a given location is often unavailable~\cite{GethCIRED2023, nyserda}, and/or \textit{2)} only sequence component impedance information is available, which is insufficient to retrieve the phase-space impedance matrices needed in untransposed multiconductor systems like DNs~\cite{kersting2018distribution}. Furthermore, overhead lines with the same (known) wire thickness and material, might still result in different (unknown) impedance matrices due to varying distances between the wires, which   depend on the line construction/installation.
That is, the input required for Carson's equations is often simply not available (to sufficient degree), and the impedances used in present utility datasets are approximate estimates~\cite{GethCIRED2023}.  

Inaccurate impedance values have a detrimental impact on power systems calculations, e.g., (optimal) power flow~\cite{Cherot2023}.
These may have been negligible with the fit-and-forget DN management approach in the past, but they can lead to seriously problematic planning and operations decisions in modern DNs, in the light of electrification, distributed generation and active DN management. Thus, research efforts have been dedicated to the data-driven derivation of impedances, exploiting the increasing digital telemetry, e.g., smart meters (SMs). Data-driven estimation appears to be the only practical alternative where inverter probing~\cite{Cavraro2019CNS} is not possible. 

This paper proposes a state estimation-based approach for impedance estimation. The process is not ``purely" data-driven, but strongly embeds system domain knowledge, making the proposed method distinct from prior work.

\subsection{Related Work on Data-Driven Impedance Estimation}

In our previous work~\cite{Vanin2023IE}, we observed that most of the prior impedance estimation (IE) methods are hardly applicable to generic
real-world DNs, due to the following:
\begin{enumerate}
    \item they consider balanced (positive-sequence) impedance models~\cite{Zhang2021, Zhang2020Topology, Guo2022,Marulli2021}\footnote{We note that~\cite{Marulli2021} does consider a three-phase system, but treats each phase separately for the impedance estimation process, without addressing mutual coupling explicitly.}, \cite{Kapoor2024, Sang2024pscc, Mittal2024}, whereas DNs are unbalanced multiconductor systems,
    \item they rely on synchrophasors~\cite{Li2022, GuptaPaolone, Moffat2020}, which are not available \textit{at scale}, particularly in low voltage DNs, 
    \item they assume noiseless measurements~\cite{ClaeysCIRED2021, Guo2022},
    \item they require a larger amount of measurement devices than realistic, especially for low voltage DNs~\cite{Dutta, WangPMAPS2020},
    \item they require extensive datasets, i.e., very long measurement time-series~\cite{Jaepil2022, Peppanen2016, Yang2022}.
\end{enumerate}

For completeness, we note that \textit{1)} impedance values might also be inaccurate in transmission networks~\cite{Costa2022}, in which case balanced (positive-sequence) IE is justified, and that \textit{2)} specific IE methods have been developed for North-American split-phase circuits~\cite{Short2013,Lave2019}. An additional difference between transmission and distribution networks is that the former usually only have a few untrustworthy parameters that need to be estimated, in an otherwise well-known network~\cite{Lin2018framework, bible}. Occasionally, this assumption is extended to DNs~\cite{Alam2024}, but it does not hold in general: all branch impedances are usually inaccurate, calling for different IE methods.

Dutta et al.~\cite{Dutta} assume that measurements are available at all buses in the network. If this is the case, unique impedance values can be retrieved. However, measurement devices in DNs are usually only placed at user/load buses~\cite{Li2022, Moffat2020}. IE may still be possible, but the solution is not unique (this is also shown in~\cite{9858017,Li2019}, where non-monitored buses are called ``hidden nodes''). Nevertheless, both~\cite{Vanin2023IE, Brouillon2024}, as well as this paper, find that workable equivalent impedance models for power system analysis can be obtained. Broullion et al.~\cite{Brouillon2024} additionally carry out a theoretical analysis on the biases that can be expected without synchrophasors for different impedance and admittance estimation methods, showing that estimating impedances is preferable.

Papers that fall into drawback \textit{5)} typically do so because of abstractions or approximations of the circuit physics, instead of preserving the non-convex relationship between power, voltage and current.  
For instance, ~\cite{Peppanen2016, Cunha2020} perform a linear regression, which adds noise in form of modelling error to the (inevitable) measurement noise~\cite{Kapoor2024}, requiring larger amounts of data to be ``filtered". Ban et al.~\cite{Jaepil2022} and Yang et. al.~\cite{Yang2022} abstract the network physics even further by resorting to data-intense machine learning methods.

\begin{table*}[t]
\vspace*{-\baselineskip}
\caption{Multi-conductor series impedance estimation and calculation literature. $z_d, z_0$ indicate direct and zero sequence components.}
\centering
\scalebox{0.85}{
\begin{tabular}{l|l|l|l}
\hline
\textbf{Category} & \textbf{Input} & \textbf{Output} & \textbf{References}  \\ \hline
Calculation, standard utility practice & $z_0, z_d$ & Approximated \textbf{Z} & \cite{GethCIRED2023} \\
Calculation, forward Carson & Line/cable properties & Best-practice \textbf{Z}: \textbf{Z}$^{3 \times 3}$ or \textbf{Z}$^{4 \times 4}$ & \cite{cleenwerck2022, Kersting2011} \\
Estimation, inverse Carson & $z_0, z_d$ & Line/cable properties and best-practice \textbf{Z}$^{3 \times 3}$ or \textbf{Z}$^{4 \times 4}$ & \cite{Tam2024} \\
Estimation, without Carson & measurements & Single-phase (positive sequence) $z_d$  
& \cite{Zhang2021}-\cite{Mittal2024} \\
Estimation, without Carson & measurements & \textbf{Z}$^{3 \times 3}$ (not suitable for four-wire networks) & \cite{Vanin2023IE, Brouillon2024, Alam2024}  \\
Estimation, with Carson & measurements & \textbf{Z}$^{3 \times 3}$ or \textbf{Z}$^{4 \times 4}$ (suitable for four-wire networks) & \textbf{This paper} \\
\hline
\end{tabular}
} 
\label{tab:series_impedance_literature}
\vspace*{-1.5\baselineskip}
\end{table*}

Our previous work~\cite{Vanin2023IE} jointly performs system state and impedance estimation as a mathematical optimization problem, but does not allow to efficiently exploit system domain knowledge, resulting in a rather degenerate and computationally intensive estimation problem. Furthermore, while~\cite{Vanin2023IE} does \emph{not} fall into drawbacks \emph{1)-5)}, it estimates impedance matrices of dimension only (up to) $3 \times 3$ (\textbf{Z}$^{3 \times 3}$). 
When DNs are four-wire and their neutral is sparsely grounded, 3$\times$3 matrices, e.g., Kron's reduction of the neutral, do not capture all network physics adequately and ultimately result in unreliable power flow results~\cite{CLAEYS2022108522}.  Grounding choices differ across the globe, e.g., sparse neutral grounding is common in Europe, hence requiring 4$\times$4 impedance matrices (\textbf{Z}$^{4 \times 4}$).  
For an overview of grounding systems world-wide, refer to~\cite{Lacroix1995}.

Rayati et al.~\cite{Rayati2024pscc} propose a sequence components estimation instead of phase-space impedances, for a transposed three-phase three-wire network. However, transposition does not regularly occur in DNs, and in four-wire DNs even transposition would not result into balanced matrices like those in~\cite{Rayati2024pscc}. Cunha et al.~\cite{Cunha2020} perform impedance estimation through linear regression, exploiting the knowledge of line and cables' R/X ratio and ratios between self- and mutual impedances, in a fashion that also assumes transposition.

Finally, we note that all references in this section address the estimation of \emph{series} impedances. 
Line \emph{shunt} impedances are usually neglected both in IE and DN analysis, as their omission implies negligible accuracy loss in Low Voltage DNs (LVDNs)~\cite{Urquhart2015} like the ones addressed in this paper. Different but related work includes Ghaderi et al.~\cite{Ghaderi2023}, which estimates the grounding resistance of medium voltage cables, and Low~\cite{Low2024}, which estimates the full network admittance matrix of a DN starting from the Kron reduction of its graph. Both~\cite{Ghaderi2023} and~\cite{Low2024} rely on synchrophasors. 

Table~\ref{tab:series_impedance_literature} summarizes the existing approaches to calculate or estimate series impedance.

\subsection{Contributions}\label{s:introduction-contributions}

This paper presents a novel IE method that improves on the state of the art, including our previous work~\cite{Vanin2023IE}, by:
\begin{enumerate}
    \item supporting (up to) 4$\times$4 series impedance matrices, enabling adequate representation of generic four-wire DNs,
    \item casting impedance variables as functions of cable/line properties, through Carson's equations.
\end{enumerate}

To our knowledge, our method is the first to present these two features. 

The first feature improves the fidelity of the impedance models, making it expressive enough to capture distribution network construction practices from most of the world, including those where networks are not transposed and Kron's reduction would be a strong approximation. Moving from $3\times3$ to $4\times 4$ impedance models is not trivial. Firstly, the number of impedance variables (i.e., the sum of the entries of impedance matrices of each branch/line) increases significantly, resulting in a harder computational problem. Secondly, assumptions related to the grounding of the neutral wire need to be included. Thirdly, the modeller needs to avoid the emergence of spurious solution that are due to allowing 0~V voltages on the neutral conductor in the power flow constraints~\cite{geth2022pitfalls} that are at the basis of our state estimation-based IE method. 

The second feature requires a more sophisticated mathematical implementation with respect to methods that directly estimate impedance entries, like our previous~\cite{Vanin2023IE}, as the variable space is quite different and a large amount of extra constraints is required. However, perhaps counterintuitively, the change of variable space presents \emph{computational} and practical advantages (ease of domain knowledge introduction), as discussed in the upcoming sections, despite being more complex. In particular, the use of Carson's equation reduces the impedance estimation time by up to two orders of magnitude with respect to our previous work. 

Note that our approach generalizes to systems with fewer than four wires, and it is sufficient to drop all the terms relative to the missing conductors to address such cases. In particular, three-phase three-wire systems lack the neutral conductor, whereas single-phase segments (e.g., service cables to a single-phase end-user) are two-wire, one of which can be a neutral and can be grounded or not. 

All code and data used for this paper are made available\footnote{\textit{https://github.com/Electa-Git/ImpedanceEstimationWithCarson}}.

The rest of the paper is structured as follows: Section~\ref{s:mathematical_model} provides the mathematical background to the IE problem, Section~\ref{s:case_studies} describes the data and case studies, whose results are reported in Section~\ref{s:results}. Section~\ref{s:conclusion} concludes the paper.

\section{Mathematical Model}\label{s:mathematical_model}

Let $\timeseriesm$ be the set of time steps for which measurements are available, and let $\measm$ be the set of such measurements.
 Impedance estimation is cast as a mathematical optimization (maximum likelihood) problem, where the objective function minimizes the sum of measurement residuals $\rho_{m,t}$:
\begin{equation}
\text{minimize} \; \;    \sum_{\substack{m \in \measm}} \sum_{\substack{t \in \timeseriesm}}   \rho_{m, t}.         \label{eq:objective} 
\end{equation}

The objective is similar to that of conventional state estimation, except that a time dimension is added. We opt for a (exact relaxation of) weighted least absolute value minimization, as opposed to the more conventional weighted least squares\footnote{This choice is motivated in our previous work~\cite{Vanin2023IE}, and has to do with the bad data rejection capabilities of the least absolute values.}:
\begin{eqnarray}
     \rho_{m, t} \geq \frac{x_{m,t} - z_{m,t}}{\sigma_m},\quad \forall m \in \measm, t \in \timeseriesm \label{eq:rWLAV1},\\
     \rho_{m, t} \geq - \frac{x_{m, t} - z_{m, t}}{\sigma_m},\quad  \forall m \in \measm, t \in \timeseriesm. \label{eq:rWLAV2}
\end{eqnarray}
The $x_{m,t}$ in \eqref{eq:rWLAV1}-\eqref{eq:rWLAV2} are state variables, and $z_{m,t}$ their corresponding measurement. The measurement's weight, $1/\sigma_m$, indicates the confidence on the quality of measurement $m$, depending on the SM accuracy class. We assume that the only available measurements are fifteen minute averages of active power ($P$), reactive power ($Q$), and (phase-to-neutral) voltage magnitude ($U^\text{mag,pn}$), as is typical for European SMs.

As any other IE  method, we assume that the network topology, including phase connectivity, is a known input. While topological information is known to be error-prone~\cite{GethCIRED2023}, reliable data-driven methods have been developed that can derive topology and/or phase connections from smart meter data, e.g.,~\cite{DekaTutorial, Vanin2024PI}, prior to performing IE.

As such, we can describe a DN as a graph with branches (edges) \branches \ and nodes \buses. The proposed method applies to both radial and meshed networks. Let $\mathcal{U}$ be the set of connected users (representing load/generation/storage/...) with a SM. Every branch $l \in \branchesm$ connects two nodes $i \in \busesm$ and $j \in \busesm$, such that the connectivity mapping is $lij \in \mathcal{C} \subseteq  \branchesm \times \busesm \times \busesm$. 
Every user $u \in \mathcal{U}$ is connected to a bus~$i \in \busesm$ and the connectivity mapping is $ui \in \mathcal{C}^{\text{u}} \subseteq  \mathcal{U} \times \busesm$.

A bold typeface indicates vectors and matrices. 
Vectors are used to stack conductor values, i.e., for a generic  phasor quantity: $\mathbf{v} = [v_p]_{\forall p \in \mathcal{P}}$, where $\mathcal{P} \subseteq \{a,b,c,n\}$ are the conductors that interest the quantity described by \textbf{v}.
$\mathbf{U}^\text{re}_\text{i,t}+\text{j} \mathbf{U}^\text{im}_\text{i,t}$ is the complex voltage phasor at bus $i$, time $t$, $\crmut+\text{j} \cimut$ is the complex current injection of user $u$ at time $t$; $\crmlijt+\text{j} \cimlijt$ is the complex current in line $l$ in the direction of $i$ to $j$ at time $t$.

The explicit neutral representation of power flow equations in power-voltage variables is known to lead to spurious solutions~\cite{geth2022pitfalls}. To avoid this issue, we use the current-voltage variable space (in rectangular coordinates, i.e., the so-called multiconductor IVR power flow formulation). 

To project the measurements $z_m$ of choice to the variable space of the IVR, the following equality constraints are enforced for every \emph{measured} user:
\begin{eqnarray}\label{eq:square}
     &  \text{U}^{\text{mag}}_\text{i,t,pn}  = \sqrt{(\text{U}^{\text{re}}_{\text{i,t,p}} - \text{U}^{\text{re}}_{\text{i,t,n}})^2 + (\text{U}^{\text{im}}_{\text{i,t,p}} - \text{U}^{\text{im}}_{\text{i,t,n}})^2 } \nonumber \\
     &
     \forall i  \in \busesm, \forall t \in \timeseriesm, \forall p \in \mathcal{P} \setminus \{n\},\label{eq:vm_meas} \\
     & \text{P}_\text{u,t,p} = (\text{U}^{\text{re}}_{\text{i,t,p}} - \text{U}^{\text{re}}_{\text{i,t,n}}) \cdot \text{I}^{\text{re}}_{\text{u,t,p}} + (\text{U}^{\text{im}}_{\text{i,t,p}} - \text{U}^{\text{im}}_{\text{i,t,n}}) \cdot \text{I}^{\text{im}}_{\text{u,t,p}} \nonumber \\
     & \forall ui \in \mathcal{C}^{\text{u}}, \forall t \in \timeseriesm, \forall p \in \mathcal{P} \setminus \{n\},\label{eq:p_meas}\\
          & \text{Q}_\text{u,t,p} = (\text{U}^{\text{im}}_{\text{i,t,p}} - \text{U}^{\text{im}}_{\text{i,t,n}}) \cdot \text{I}^{\text{re}}_{\text{u,t,p}} - (\text{U}^{\text{re}}_{\text{i,t,p}} - \text{U}^{\text{re}}_{\text{i,t,n}}) \cdot \text{I}^{\text{im}}_{\text{u,t,p}} \nonumber \\
     & \forall ui \in \mathcal{C}^{\text{u}}, \forall t \in \timeseriesm, \forall p \in \mathcal{P} \setminus \{n\},\label{eq:q_meas}
\end{eqnarray}
where the left-hand sides correspond to the $x_{m,t}$ in \eqref{eq:rWLAV1}-\eqref{eq:rWLAV2}. Note that while SMs measure phase-to-neutral voltages, the state phasors $\urit+\text{j} \uiit$ are defined with respect to the ground. 
If the neutral is perfectly grounded at all buses, Kron's reduction can be used, and ground and neutral voltages can be assumed equal, which would simplify the equations above. However, this does not hold in general~\cite{CLAEYS2022108522}.

\begin{figure}[b]
    \centering
\includegraphics[width = 0.45\textwidth]{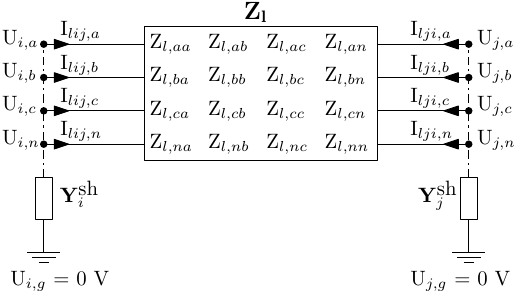}
    \caption{Four-wire series impedance and neutral-ground shunt model.}
    \label{fig:four-wire-impedance}
\end{figure}

The multi-conductor Ohm's law is:
\begin{eqnarray}\label{eq:ohm_vcr}
    \urit = \urjt + \mathbf{R}_{\text{l}}  \crmlijt - \mathbf{X}_{\text{l}} \cimlijt,\; \forall lij \in \mathcal{C}, t  \in \timeseriesm, \\
    \uiit = \uijt + \mathbf{R}_{\text{l}} \cimlijt + \mathbf{X}_{\text{l}}  \crmlijt, \; \forall lij \in \mathcal{C}, t \in \timeseriesm. \label{eq:ohm_vcr2}
\end{eqnarray}

Measurements and electric states (voltages, currents, etc.) are time-variant, whereas impedances ($\mathbf{Z}_\text{l} = \mathbf{R}_{\text{l}} + \imagunit \mathbf{X}_{\text{l}}$) are time-invariant. In conventional DSSE impedances are known inputs, whereas here the impedance matrix entries are estimated jointly to the time-dependent states. 

The multi-conductor Kirchhoff's current law is:
\begin{eqnarray}\label{eq:kcl}
   & \sum_{lij \in \mathcal{C}} \! \crmlijt + \!\! \sum_{ui \in \mathcal{C}^{\text{u}}} \!\crmut + \mathbf{Y}_\text{i}^{\text{sh}} \circ \urit = 0, \, \forall t \in \timeseriesm,  i \in \busesm, \\ 
    & \sum_{lij \in \mathcal{C}}\!
    \cimlijt + \sum_{ui \in \mathcal{C}^{\text{u}}} \cimut + \mathbf{Y}_\text{i}^{\text{sh}} \circ \uiit = 0, 
   \, \forall t \in \timeseriesm,  i \in \busesm,
\end{eqnarray}
where $\circ$ indicates element-wise multiplication and $\mathbf{Y}_\text{i}^{\text{sh}}$ is the nodal shunt admittance of bus $i$. In this paper, we assume that, except for the transformer's bus, which is perfectly grounded, $\mathbf{Y}_\text{i}^{\text{sh}} = 0 \, \, \forall i$. This is a generally considered an accurate representation of European LVDNs~\cite{CLAEYS2022108522}. Non-zero shunt admittances are straightforward to add as either input, if known, or as variables. In the latter case, they can be estimated with the method proposed in this paper. However, for conciseness, we leave this for future work: a more extensive discussion of grounding practices would be required. 

We note that distribution networks are characterized by a large number of zero-injection buses. With the present implementation, these are implicitly included as equality constraints, as common to multiple state-of-the-art DSSE implementations~\cite{Vanin2022}, to avoid the ill-conditioning that would otherwise be caused by the use of low virtual measurement weights.

The impedance model for a four-wire branch is illustrated in Fig.~\ref{fig:four-wire-impedance}. While different combinations are possible, in this work we assume that the branches are either three-phase four-wire (``main" portions of the feeder), or single-phase two-wire (service cables). This is the standard construction approach for LVDNs, a.o., in Europe and Australia. As such, we can write $\branchesm = \branchestpm \cup \branchesspm$. 

Two-wire branches are described by three series impedance elements, $Z_{l, pp}, Z_{l, pn}, Z_{l, nn}$. 
 Consequently, the sizes of current and power vectors are also halved. 

\begin{figure}[t]
    \centering
\includegraphics[width = 0.35\columnwidth]{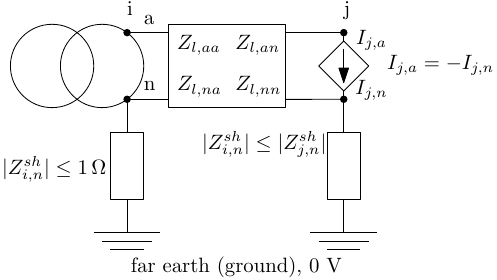}
    \caption{Two-bus single-phase example of component model with explicit neutral and grounding. Note that imperfect, non-zero grounding is possible.}
    \label{fig:grounding}
\end{figure}

Note that line impedance matrices always have \textbf{symmetric} real and imaginary parts by construction, i.e.: 
\begin{equation}\label{eq:symmetry}
    Z_{l, pq} = Z_{l,qp} \quad \forall l \in \branchesm, p,q \in \mathcal{P}.
\end{equation}

Finally, Fig.~\ref{fig:grounding} illustrates the electrical component model that complements the problem formulation. The neutral and phase currents are the same at load/generator buses, but the return line flows depend on the neutral grounding (again, in this paper $|Z^{sh}| \equiv (\textbf{Y}^{sh})^{-1} = 0$).

\subsection{Use of construction codes}

\begin{figure}[t]
    \centering
\includegraphics[width = 0.53\textwidth]{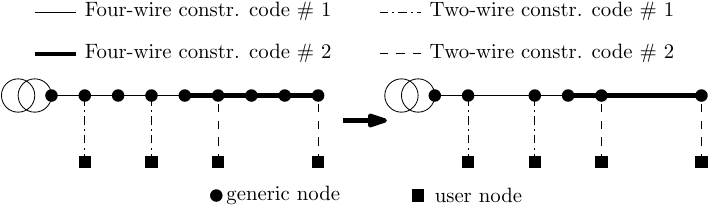}
    \caption{Illustration of bus reduction and construction codes.}
    \label{fig:bus-reduction}
\end{figure}

In our previous work, and the rest of IE literature, variables $R_{l, pq}, X_{l, pq}$ are ``learned" directly. In this paper, on the other hand, they are dependent on variables that characterize their construction code. Construction codes collect information on conductor properties (material, cross-section, etc.) and - possibly - their layout (e.g., how far away they are placed from each other in a overhead line), and are the way network operators log line and cable installation data. Segments of the DN with the same construction code $k \in \mathcal{K}$ are characterized by the same nominal ($\Omega/km$) impedance, which operators then multiply by the segments' length, which is estimated from GIS data. The number of construction codes deployed in a specific DN is generally much smaller than that of the branches, as illustrated in Fig.~\ref{fig:bus-reduction}, where two different four-wire and two-wire construction codes are used. This occurs when a first portion of the feeder is built at one time, and subsequently extended - at a time where the previously used cable types are no longer available or preferred. Fig.~\ref{fig:bus-reduction} additionally illustrates a reduction of the number of nodes in the DN topology (from the left to the right of the figure) that, \emph{without approximation}, reduces the number of the equivalent IE solutions, and is beneficial for the optimizer. This reduction was used in our previous work~\cite{Vanin2023IE}.

\begin{figure}[t]
    \centering
\includegraphics[width = 0.35\textwidth]{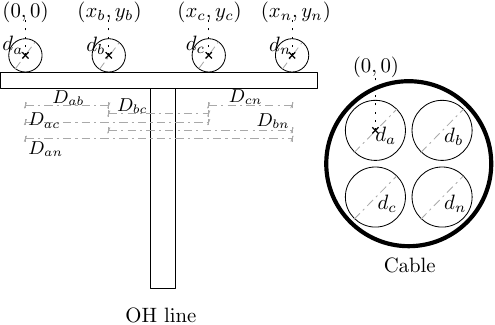}
    \caption{Four-wire line and cable geometrical properties. Cable distances and coordinates, omitted in the figure, are derived similarly to the OH line. The illustrated OH line layout example is common in Australia.}
    \label{fig:four-wire-geometry}
\end{figure}

Fig.~\ref{fig:four-wire-geometry} illustrates the geometry-related variables for two generic four-wire construction codes corresponding to a overhead (OH) line and a cable. Every branch of a DN is assigned a construction code, with a given nominal impedance $\mathbf{Z}^{\text{nom}}_{kl}$: 
\begin{equation}\label{eq:zl}
\mathbf{Z}_l~=~\mathbf{Z}^{\text{nom}}_{kl}\cdot~\ell_l~\, \,~\forall~kl~\in~\branchesm^k,
\end{equation}
where $\branchesm^k \subset \branchesm$ is the set of branches of construction code $k$. Note that $\mathbf{Z}^{\text{nom}}_{kl}$ is the same for all branches with construction code $k$, whereas the length $\ell_l$ is branch-specific.

The use of construction codes (a.k.a. line codes) has two advantages. Firstly, fewer variables are needed than methods that derive an independent impedance matrix for every branch, as discussed in Section~\ref{s:number-of-variables}. This is because branches with the same code share the same nominal impedances. Secondly, a better exploitation of domain knowledge is possible, as discussed in Section~\ref{s:domain-knowledge}. The latter stems from the fact that conductor manufacturers and installation practices are a finite set and utilities may know them well. As such, it is possible to bind the variables to reasonable values.

These advantages mitigate degeneracy and search space sizes and issues, resulting in shorter solve times, lower risk of overfitting, and better variance-bias trade-offs. The above, together with the fact that we are the first to use construction codes in this sense, summarizes our contribution \emph{3)}.

Mathematically, construction codes can be translated into impedance matrices through Carson's equations.

\begin{table*}[b]
\caption{\# variables with different impedance estimation methods - the total refers to the reduced (right-hand) feeder in Fig.~\ref{fig:bus-reduction}.}\label{tab:variables-fig}
\scalebox{0.8}{
\begin{tabular}{l|c|c|c|c|c|c}
\hline
Method & \# $\ell$ & \# $\mathbf{R}$ var. & \# $\mathbf{X}$ var. & \# $\mathbf{\bar{R}}$ var. & \# $\mathbf{\bar{X}}$ var. & Tot. Fig.~\ref{fig:bus-reduction}\\ \hline 
LLE & $|\branchesm|$ & - & - & - & - & 9 \\
IME in~\cite{Vanin2023IE} (up to 3$\times$3 matr.) & - & 6$\cdot |\branchestpm|$ + 1$\cdot |\branchesspm|$  & 6$\cdot |\branchestpm|$ + 1$\cdot |\branchesspm|$ & - & - & 68 \\
IME in~\cite{Vanin2023IE}, transp. R. & - & 4$\cdot |\branchestpm|$ + 1$\cdot |\branchesspm|$  & 6$\cdot |\branchestpm|$ + 1$\cdot |\branchesspm|$ & - & - & 58 \\
Extension of~\cite{Vanin2023IE} (up to 4$\times$4) & - & 10$\cdot |\branchestpm|$ + 3$\cdot |\branchesspm|$  & 10$\cdot |\branchestpm|$ + 3$\cdot |\branchesspm|$ & - & - & 116 \\
Extension of~\cite{Vanin2023IE}, transp. R & - & 5$\cdot |\branchestpm|$ + 3$\cdot |\branchesspm|$  & 10$\cdot |\branchestpm|$ + 3$\cdot |\branchesspm|$ & - & - & 91 \\
This paper (up to 4$\times$4) & $|\branchesm|$  &  - & - & 4$\cdot |\mathcal{K}^{\text{3ph}}|$ + 2$\cdot |\mathcal{K}^{\text{1ph}}|$ & 10$\cdot |\mathcal{K}^{\text{3ph}}|$ + 3$\cdot |\mathcal{K}^{\text{1ph}}|$  & 47 \\
Worst case with Carson & $|\branchesm|$  &  - & - & 4$\cdot |\branchestpm|$ + 2$\cdot |\branchesspm|$ & 10$\cdot |\branchestpm|$ + 3$\cdot |\branchesspm|$  & 90 \\
\hline
\end{tabular}
} 
\end{table*}

\subsection{Carson's Equations}

 Carson's equations are usually presented for a 60~Hz power system, in Imperial units (North American case). Here, we use the 50~Hz, International system version of Carson (Australian and European case), derived by Cleenwerck et al.~\cite{cleenwerck2022}: 
\begin{eqnarray}\label{eq:Zij}
    & \Ziibar = \Riibar + \imagunit \Xiibar \quad \forall k \in \mathcal{K}, \; \; \left[ \frac{\Omega}{km}\right] \nonumber \\
    & \Zijbar = \Rijbar + \imagunit \Xijbar \quad \forall k \in \mathcal{K}, \; \; \left[ \frac{\Omega}{km}\right] \nonumber
\end{eqnarray}
where:
\begin{eqnarray}\label{eq:RXCleenwerck}
  \Riibar & = & \ri+0.049348, \label{eq:Rpp} \\
 \Rijbar & = & 0.049348, \label{eq:Rpq} \\
  \Xiibar & = & 0.062832\left( ln\left( \frac{1}{\coefone \gmri}  \right) + \coeftwo\right), \\
  \Xijbar & = & 0.062832\left( ln\left( \frac{1}{\coefone \dij}  \right)  + \coeftwo \right).
\end{eqnarray}

\begin{itemize}
    \item $\ri$ is the conductor's ac self-resistance [$\Omega/km$]
    \item $\gmri$ is the geometric mean radius of conductor $p$  [$mm$]
    \item $\dij$  is the distance between conductor $p$ and $q$ [$mm$]
    \item $\coefone = 3.28084 \cdot 10^{-3}$ and $\coeftwo $ = 8.0252 are conversion constants to change the system of units [-]
\end{itemize}
$\dij > 0$ always holds, ensuring that the logarithm of its inverse is defined. Distances between conductors $\dij$ can be defined from the $(x,y)$ coordinates of their centers\footnote{One conductor is chosen as reference and assigned coordinates (0,0).}:
\begin{equation}\label{eq:dist}
    (\dij)^2 =   (\xxi- \xj)^2 + (\yi - \yj)^2 \, \, \, \text{[mm}^2\text{]}   .
\end{equation}
For a round conductor, the $\gmri$ is,
\begin{equation}\label{eq:gmr}
    \gmri = exp(-\relperm/4)\cdot \sqrt{\frac{\Ai}{\pi}} \; \; \forall \idcondone \in \setCond,
\end{equation}
where $\Ai$ is the conductor cross-section, and $\relperm$ is the relative permeability of the conductor material. The latter is set to 1, as is the case for all common conductor materials~\cite{cleenwerck2022}.

Let $d_{p}$ be the diameter of a conductor, then the area is,
$$ A_p = \pi d_p^2/4.$$

For cables that are composed by a set of strands $\mathcal{S}$, with cardinality $|\mathcal{S}| = S$, the following can be used instead~\cite{cleenwerck2022},
\begin{equation}\label{eq:gmr-strands}
    \gmri = \sqrt[S]{\prod_{i \in \mathcal{S}} \left(e^{-\relperm/4}\cdot \sqrt{\frac{A_i}{\pi}} \prod_{i \in \mathcal{S}, i \neq j} D^{str}_{ij} \right)^{\frac{1}{S}}} ,
\end{equation}

where $D^{str}_{ij}$ is the distance between the strands.

Finally, the ac self-resistance is,
\begin{equation}
    \ri =  \frac{\rhoi}{\Ai}(1+(\alpha_{p,k} (T-20 ^\circ \text{C}))), \label{eq_self_ac}
\end{equation}

where $T$ [$^{\circ}$C] is the ambient temperature, $\rhoi$ is the (material-dependent) conductor resistivity, and $\alpha_{p, k}$ is the temperature coefficient of the material.
We assume proximity and skin effect are negligible, which Urquhart and Thomson show to be an acceptable approximation in LVDNs~\cite{Urquhart2015}.

\subsection{Number of Variables: Comparison with Other Approaches}\label{s:number-of-variables}

It is interesting to observe the difference between the proposed method and other methods in the literature. Some authors perform line length estimation (LLE), assuming that the $\mathbf{Z}^{\text{nom}}_{kl}$ of each branch is known, and only branch lengths $\ell$ need to be estimated~\cite{ClaeysCIRED2021, Vanin2023IE}. This is generally not realistic, but results in $|\branchesm|$ variables only.

With (up to) 3$\times$3 impedance matrix estimation (IME) in~\cite{Vanin2023IE}, branch length variables are not used. For every single-phase branch, one resistance and one reactance variable suffice, whereas for a three-phase branch, twelve variables are needed (as $R$ and $X$ are symmetrical, only their upper or lower triangular elements are estimated). Extending such IME (without Carson) to explicitly include the neutral would require 20 and 6 variables for three- and single-phase branches, respectively.

With Carson's equations, on the other hand, the variables consist of a $\ell_l$ for each branch $l$ and $\mathbf{Z}^{\text{nom}}_{kl}$ for every linecode $k$. 
 Considering single- and three-phase line codes, such that $\mathcal{K}~=~ \mathcal{K}^{\text{1ph}}~\cup~\mathcal{K}^{\text{3ph}}$, the number of variables is 14 for three-phase codes, and 5 for single-phase codes. Note that the off-diagonal resistances in~\eqref{eq:Rpq} are constants.  

Table~\ref{tab:variables-fig} reports the number of series impedance-related variables for the reduced feeder in Fig.~\ref{fig:bus-reduction}. Although not tested in this paper, and not discussed in other literature, the table also illustrates the number of variables for a hypothetical estimation in which R and X values are learned directly (similar to~\cite{Vanin2023IE}), but where all off-diagonal R values are the same. This would be a ``hybrid" case, in which domain knowledge is enforced only in that - like with Carson's equations - we assume that the off-diagonal resistance values are only a function of the ground's contribution. These are indicated in the table with a ``transp. R" for both a three- and four-wire case. It can be observed how the number of variables using Carson is significantly lower with respect to the IME case, even under the ``transp. R" assumption. The worst case scenario with Carson (last row in the table) occurs if there is no prior information in the branch-construction code assignment. In this case, each branch can be assigned a unique line code, and still leads to fewer variables than the $4\times 4$ IME.

Finally, note that in addition to the impedance-related variables, the system's states are also treated as variables (as our IE is based on DSSE). The states correspond the variable space of the (nonlinear, non-convex) IVR formulation, as described in the beginning of this section. Four-wire models, as such, not only have more impedance variables, but more ``states", too, compared to three-wire models.

\subsection{Exploiting Domain Knowledge}\label{s:domain-knowledge}

In the estimation problem thus far \eqref{eq:objective}-\eqref{eq_self_ac}, there is no restriction on the values of cable coordinates/distances and cross-sections. Variable bounds have been added, based on common sense and real-world practices, e.g., $D_{k,pq}~\leq~100~mm$ is a more than safe assumption for LV cables.\footnote{Due to space reasons, please refer to the repository's \emph{src/core/variable.jl} file for the bounds' values we used here.} No additional constraints are added. We call this the \emph{non-restricted case}, labelled as \texttt{No rest.} in tables and figures.

Hereafter, we define three ways to exploit such knowledge, defined by different constraints. Such constraints can be applied either to all construction codes, or a subset. In this paper, they are applied to all, so \textit{we neglect the $k$ subscripts}. 

\subsubsection{Cross-section restrictions}

In cables, the phase conductors are generally the same size, whereas the neutral can be up to 50\% smaller in cross-section. Thus, we can enforce,
\begin{eqnarray}
   & A_p = A_q \, \, \forall p,q \in \mathcal{P}\setminus\{n\}, \label{eq:Apq_ug} \\
  & 0.5 A_p \leq A_n \leq A_p. \label{eq:Apn_ug}
\end{eqnarray}
This set of constraints is abbreviated as $\texttt{A}_\texttt{p}$ \texttt{rest.}

\subsubsection{Layout restrictions}

Cable geometry is generally that of Fig.~\ref{fig:four-wire-geometry}, so we can assume $ y_a = y_b, \quad x_a = x_c$, which implies:
\begin{eqnarray}
    D_{ab} = D_{ac} = \frac{D_{bc}}{\sqrt{2}}, \, \, \, \, 
    D_{an} = \frac{D_{ab}}{\sqrt{2}}+D_{ac} , \label{eq:Dpn_ug1} \\
    D_{bn} = D_{cn} = \sqrt{D_{ac}^2+\frac{D^2_{ab}}{2}}\label{eq:Dpn_ug2}  .
\end{eqnarray}
The use of these constraints is indicated as  \texttt{$\mathcal{G} \, \, rest.$} Eq.~\eqref{eq:Dpn_ug1}-\eqref{eq:Dpn_ug2} allow the neutral to be smaller than phase conductors.
 
\begin{table*}[b]
\caption{Summary of equations used and independent impedance variables by different exploitation of domain knowledge. $A_p$ is a generic cross-section variable when all phase conductors have the same size.}
\centering
\begin{tabular}{l l l }
\hline
\textbf{Name} & \textbf{Impedance Variables} & \textbf{Mathematical model} \\ 
\hline
\texttt{No rest.} & $\ell, A_a, A_b, A_c, A_n, x_b, x_c, x_n, y_b, y_c, y_n$ & \eqref{eq:objective}-\eqref{eq:zl} \\
$\texttt{A}_\texttt{p}$ \texttt{rest.} & $\ell, A_p, A_n, x_b, x_c, x_n, y_b, y_c, y_n$ & \eqref{eq:objective}-\eqref{eq:zl}, \eqref{eq:Apq_ug}-\eqref{eq:Apn_ug} \\
\texttt{$\mathcal{G}$ rest.}  & $\ell, A_a, A_b, A_c, A_n, D_{ab}, D_{an}, D_{cn}$ & \eqref{eq:objective}-\eqref{eq:zl}, 
\eqref{eq:Dpn_ug1}-\eqref{eq:Dpn_ug2} \\
\texttt{$\mathcal{G}$+}$\texttt{A}_\texttt{p}$ \texttt{rest.} & $\ell, A_p, A_n, D_{ab}, D_{an}, D_{cn}$ & \eqref{eq:objective}-\eqref{eq:zl}, \eqref{eq:Apq_ug}-\eqref{eq:Apn_ug}, 
\eqref{eq:Dpn_ug1}-\eqref{eq:Dpn_ug2}  \\ \hline
\end{tabular}
\label{tab:domain-knowledge}
\end{table*}

\subsubsection{Combination of the above}

Abbreviated as \texttt{$\mathcal{G}$+}~$\texttt{A}_\texttt{p}$~\texttt{rest.}, employs both sets of constraints above.

Note that the domain knowledge exploitation does not affect \eqref{eq:objective}-\eqref{eq:symmetry}; these equations are employed as-is at all times.

\subsection{General Assumptions and Multiple Solutions}

\begin{figure}[t]
    \centering
\includegraphics[width = 0.45\textwidth]{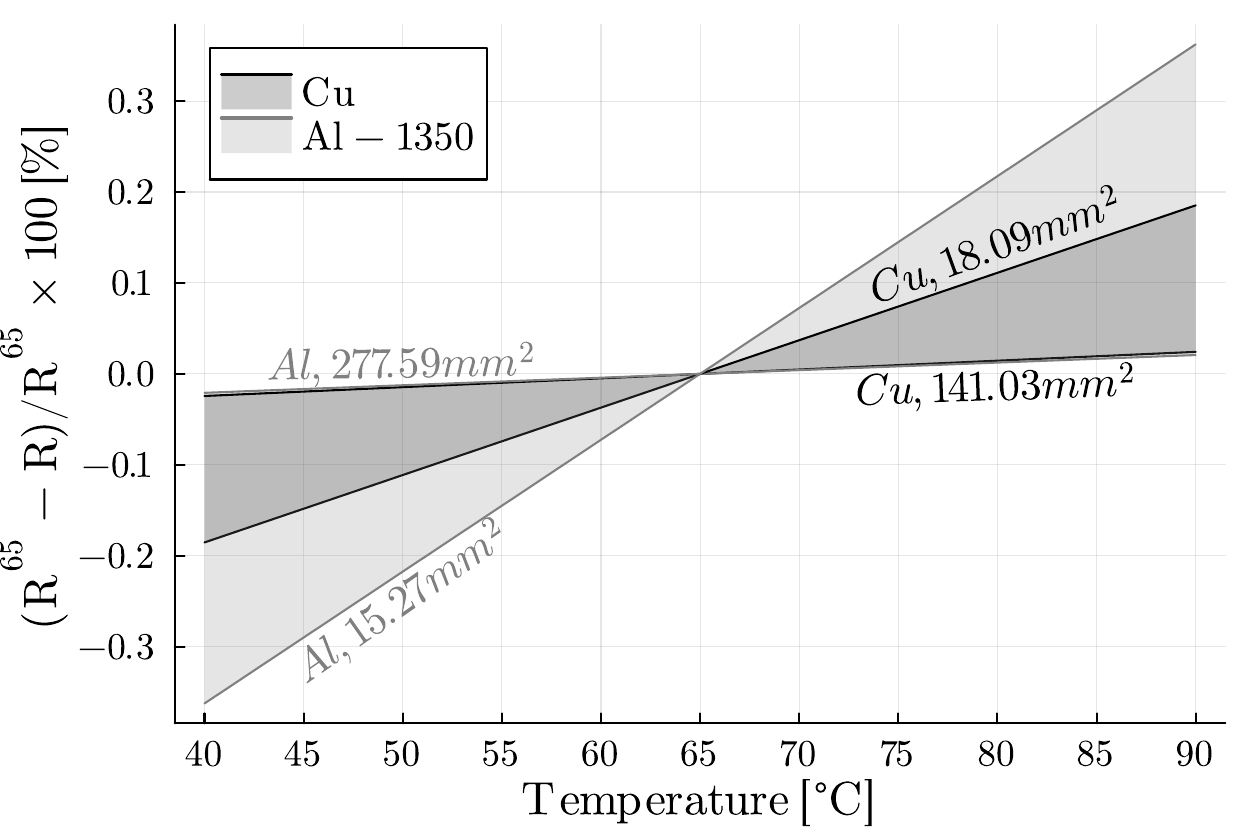}
    \caption{Self resistance errors with respect to assuming T=65. Low resistivity and high cross sections imply lower errors. Colored areas span common cross-section sizes of copper and aluminium conductors; the aluminium area fully contains the copper's. Material properties are taken from Table V in~\cite{Tam2024}.}
    \label{fig:temperature}
\end{figure}

Throughout this paper, we make the following assumptions:
\begin{enumerate}
    \item The conductor material is known, i.e., $\rho_{p,k}, \alpha_{p,k}$ are constants. Utilities often know this, and it is often practice-dependent, e.g., LV OH lines are rarely made of copper. Furthermore, if the material is unknown, the limited possibilities (copper, aluminium) allow to perform combinatorial checks: IE is repeated fixing $\rho_{p,k}, \alpha_{p,k}$ for the different materials. The material that leads to the best result on a measurement validation set is kept. 
    \item Temperature is treated as constant. In Carson's equations \eqref{eq:Rpp}, \eqref{eq_self_ac}, temperature only affects the self resistances. Fig.~\ref{fig:temperature} illustrates the relative error in resistance values with respect to assuming 65$^\circ$C. Such error is below 0.5\% over a wide range of temperatures, which is at least one order of magnitude lower than the best-case impedance estimation accuracy (as per the numerical results in this paper as well as our industrial experience), and thus negligible. 
    \item While their geometries are unknown, we do know the number of unique construction codes in a network, and their location, i.e., which branch $l \in \branchesm$ has code $k \in \mathcal{K}$. In our experience, network operators do have reliable data on this. Should there be doubts about any branch, these can be cast as a separate construction code. If all branches are uncertain, the ``worst case with Carson" scenario is realized (see Table~\ref{tab:variables-fig}).
\end{enumerate}

Assumptions \emph{1)-2)} do not significantly reduce the computational complexity of the IME problem, which is still nonlinear, non-convex and features transcendental equations, e.g., \eqref{eq:RXCleenwerck}. In general, 
it is possible to solve the IE leaving $\rho_{p,k}, \alpha_{p,k}$ and T in as (continuous) variables. However, this would lead to a number of additional equivalent solutions. For instance, in~\eqref{eq_self_ac}, any combination of $\rho_{p,k}, A_{p,k}$ that has the same ratio, would lead to equivalent $r_{p,k}$ values. 

We believe that the combinatorial workaround for $\rho_{p,k}, \alpha_{p,k}$ is a better solution, as this might lead to finding the ``actual" (or close to) $A_{p,k}$, reducing the risk of overfitting to an unrealistic value. Exploiting domain knowledge as illustrated in the previous sections is also a method to reduce overfitting. 

The existence of multiple solutions, while in general undesired, does have an advantage: should any of the assumptions be accidentally wrong (e.g., imposing $A_{p,k} < 100~mm$ when the actual value is higher), the final $R, X$ values may still be accurate, as other variables can compensate for the bound error to provide an equivalent impedance fit. 

Multiple equivalent solutions are also present when learning $R, X$ values directly (without Carson), as discussed in~\cite{Vanin2023IE},\cite{9858017},\cite{Li2019}. As a matter of fact, we observe that binding $R, X$ entries directly, which we did in~\cite{Vanin2023IE}, is much harder and less effective (because the physical interpretation of $R,X$ entries is less intuitive and worse defined) than linking the construction variables in Carson's equations.

Finally, we assume that line lengths $\ell$ are certainly within a confidence interval, chosen here to be $\pm30\%$ from their true value. This is a rather conservative assumption, as the GIS-derived line length information utilities have is usually more accurate. Such knowledge is enforced as variable bounds: $0.7\cdot~\ell_l^{\text{true}} \leq \ell_l \leq 1.3 \cdot \ell_l^{\text{true}} \, \, \forall l \in \branchesm $. This assumption does not affect the results: because of the presence of multiple solutions, exact length values cannot be retrieved (see section~\ref{s:results}).

Table~\ref{tab:domain-knowledge} reports the independent impedance-related variables for the domain knowledge exploitation examples discussed above. These apply to the specific case/geometry of cables, but a similar derivation is possible for OH lines. 

\subsection{Summary of Prior Network Knowledge}

Topology, including phase connectivity, is required (for IE in general, not exclusive to our method). In some cases, e.g., sparsely built areas, topologies inferred from GIS data are rather accurate. When this is not true, the literature offers numerous methods for data-driven topology and phase identification, which are compatible with the SM data specifications used here. 
Even when topology and phases are known, impedance values are usually still inaccurate for a variety of reasons (see Section~\ref{s:introduction}), among other, because manufacturers only provide positive (and occasionally zero) sequence impedances, which are not sufficient to retrieve accurate models for generic untransposed networks.

However, even though the exact impedance values are not known, network operators do generally have trustworthy information on how many different construction codes are available in a feeder, and at which point these change (e.g., between main and service cables). Similarly, the material of the conductors is assumed to be known, as this is reported by the manufacturer and usually implied by the segment being overhead or underground.

The use of additional (hypothetical) domain knowledge, described in Section~\ref{s:domain-knowledge}, are also tested and their impact on the estimation process is compared.

\section{Data and Case Studies Set-up}\label{s:case_studies}
\begin{figure*}[h!]
\centering
\includegraphics[width=0.85\textwidth]{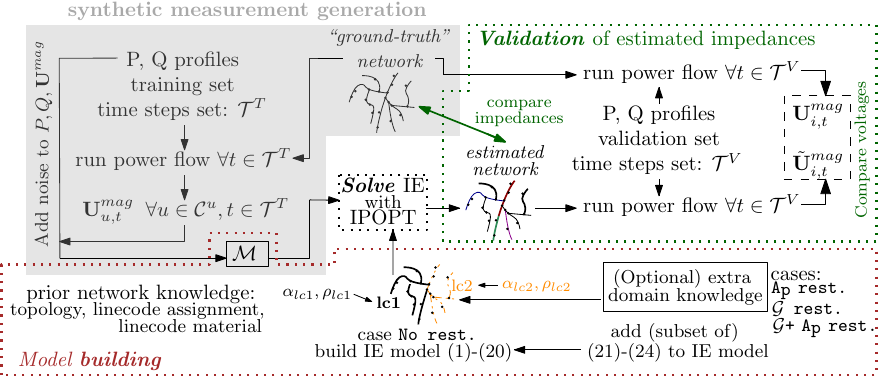}
    \caption{Illustration of the overall process. \textcolor{darkgray}{Synthetic measurements} are generated and fed into the IE model, together with appropriate a-priori knowledge of the networks' features and properties (\textcolor{brown}{model building}). Then IE is \textbf{solved} and the resulting impedance model is \textcolor{darkgreen}{validated}, both by comparing the power flow results and the estimated impedance matrix values.}
    \label{fig:flowchart}
\end{figure*}

We use the four-wire extension of two representative LV feeders from Electricity North West Limited (ENWL)~\cite{Rigoni2016}: feeder 1 network 1, i.e., the European Low Voltage Test Feeder (EU), and the thirty-load network 5 feeder 4 (30L). 

Several construction codes for underground cables employed in Queensland, Australia, curated ad-hoc for this paper, are assigned to the feeder branches. 

Table~\ref{tab:case-studies} reports an overview of the case studies. We assume that there is only one three-phase construction code, whereas service cables can be of two types. The construction code names match those in the repository.

\begin{table*}[t!]
\caption{Summary of case studies. The number of nodes is after the reduction shown in Fig.~\ref{fig:bus-reduction}.}\label{tab:case-studies}
\centering
\scalebox{0.70}{
\begin{tabular}{r|c|c|c|c|c}
\hline
\textbf{Name} & \textbf{Feeder} &  \textbf{\# Users} & \textbf{\# Nodes} & \textbf{Four-wire Construction Codes} & \textbf{Two-wire Construction Codes (service cables)}  \\ \hline
30L & ENWL ntw 5 fdr 4 - 4 wire & 30 & 58 & uglv\_240al\_xlpe &  uglv\_185al\_xlpe, ugsc\_16al\_xlpe  \\
EU & ENWL ntw 1 fdr 1 - 4 wire & 55 & 110 & uglv\_120cu\_xlpe & ugsc\_16cu\_xlpe, ugsc\_25cu\_xlpe  \\
\hline
\end{tabular}
} 
\end{table*}

The end-user profiles are a subset of the Open Energy Data Initiative's dataset  ``End-Use Load Profiles for the U.S. Building Stock"~\cite{oedi_4520}. The selected profiles consist of a year of active power values with fifteen-minute resolution, which is the same as that of SM measurements. Reactive power time series are created assigning power factors.

Synthetic measurement data is created by adding Gaussian noise to power flow inputs (P, Q) and outputs ($U^{mag,pn}$). The maximum errors (i.e., 3$\sigma$) for $U^{mag,pn}$, P and Q are, respectively, 0.5\%, 1\%, and 2\%. This mirrors the capabilities of, a.o., European SMs. This synthetic measurement creation process is standard (see, e.g.,~\cite{Vanin2023IE}).

We split the profiles in a training and a validation set, as is common practice in machine learning (ML). The time step selection greatly affects the estimation quality. In particular, time steps in which voltage variations are limited are characterized by low signal-to-noise ratio~\cite{Brouillon2024}, which make it particularly challenging to learn impedances. We observed this effect in our previous work as well. In~\cite{Vanin2023IE}, choosing the 200 time steps with higher total power flows ensured IE with acceptable quality, as power flows are a proxy for voltage drops. Developing improved time step selection for IE would be greatly beneficial\footnote{We only know of one paper investigating this, for balanced systems~\cite{Mittal2024}.}, however, for the purpose of this work, we rely on our previous results~\cite{Vanin2023IE} and pick the 200 most loaded time steps as training data. The subsequent 400 most loaded time steps are used for validation. Note that our estimation problem features (in)equality constraints and is solved through a second-order method, which generally presents improved accuracy and convergence properties with respect to the first-order ones 
usually favoured in ML.

The overall IE process: generation of synthetic measurements, IE model building, solving, and result validation, is illustrated in Fig.~\ref{fig:flowchart}.

\section{Results}\label{s:results}
The problem is implemented in \textsc{Julia} v1.9, relying on \textsc{JuMP}~v1.21.1~\cite{JuMP} and \textsc{PowerModelsDistribution} v0.15.2~\cite{PMD_PSCC}. It is solved with \textsc{Ipopt} v3.14.14~\cite{ipopt}, using the HSL's MA27 solver routine~\cite{HSL} and \textsc{MKL} v0.6.3. Calculations are run on  Intel Xeon CPU E5-4610 v4, 32 GB RAM.

The results presented here are subdivided in four groups:
\begin{enumerate}
    \item computational time, 
    \item cumulative impedance estimation accuracy,
    \item power flow validation,
    \item construction code estimation accuracy.
\end{enumerate}

Assessment of \emph{1)-3)} is possible with IE without Carson's equations (as in~\cite{Vanin2023IE}), whereas \emph{4)} is not.  

\begin{figure*}[t]
\centering
\includegraphics[width=.31\textwidth, trim={0 1.1cm 0 0},clip]{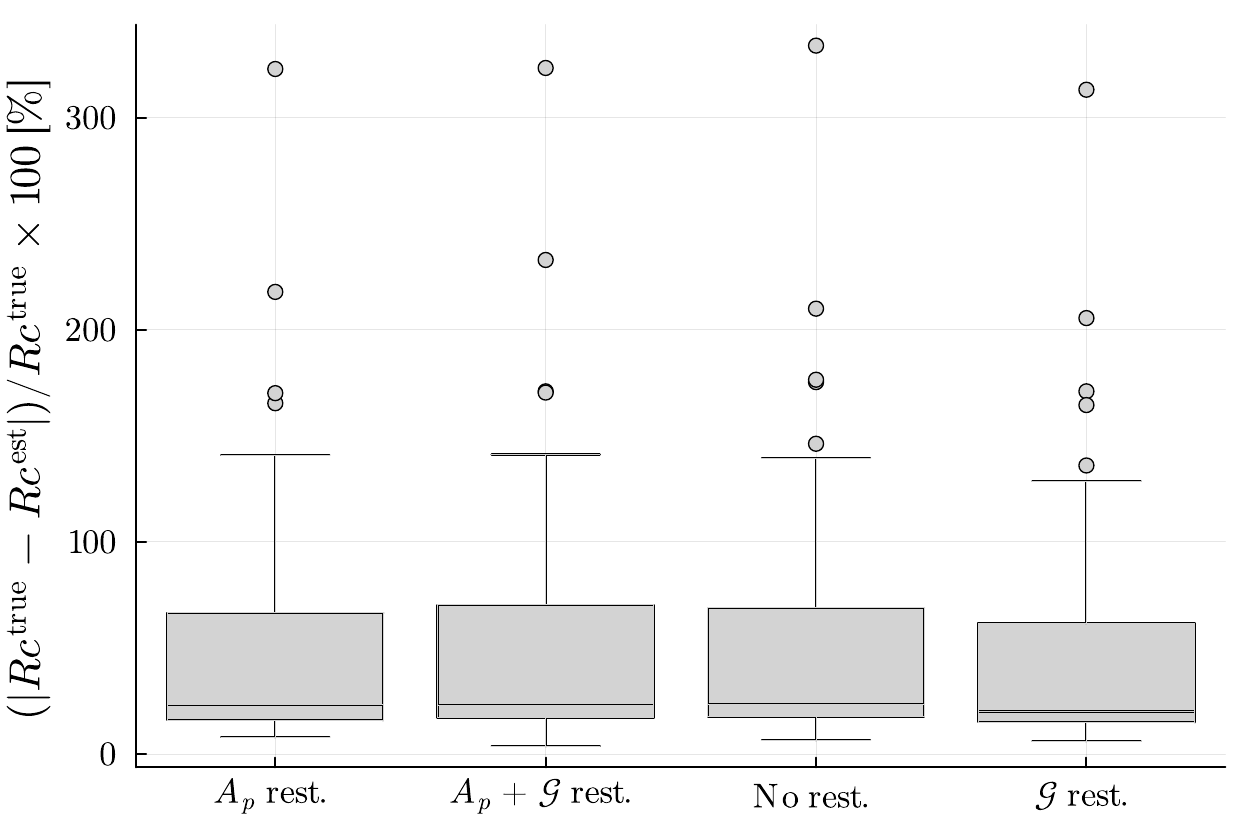}
\includegraphics[width=.31\textwidth, trim={0 1.1cm 0 0},clip]{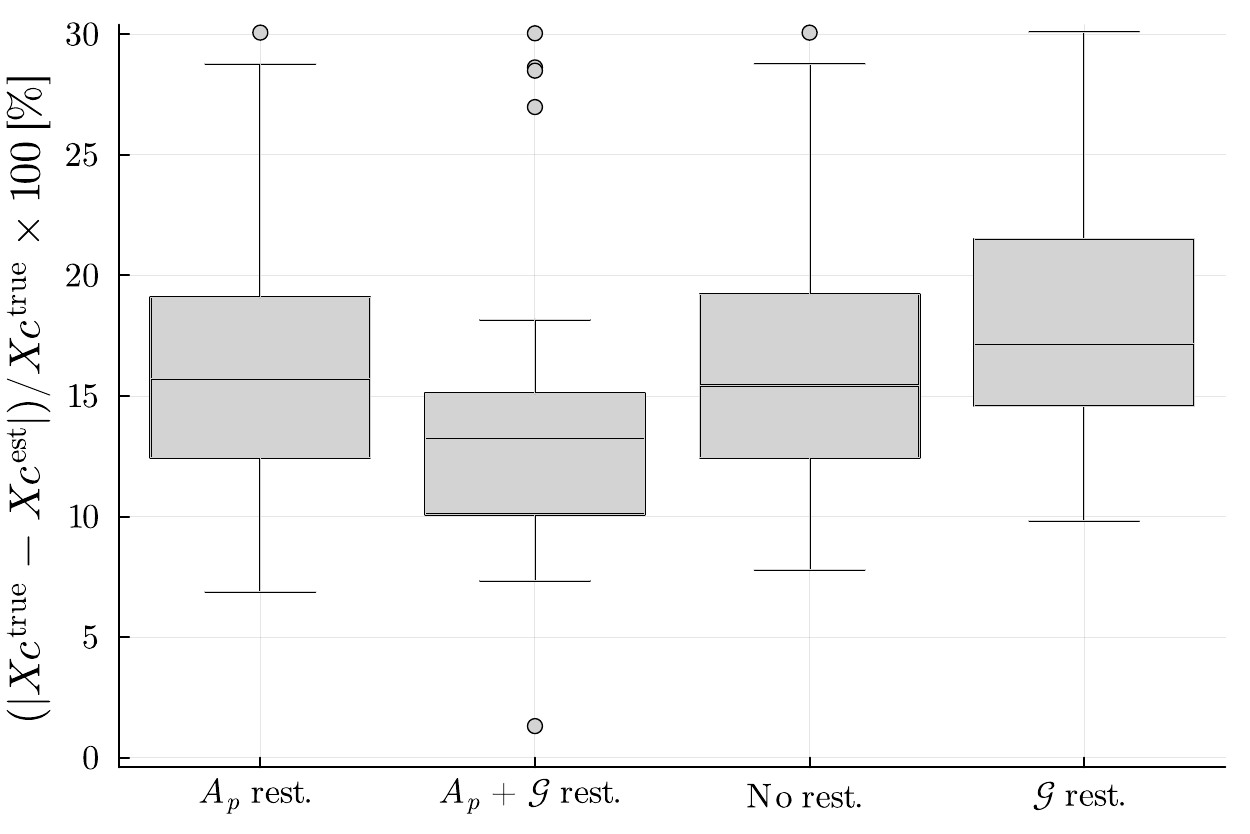}
\includegraphics[width=.31\textwidth, trim={0 1.1cm 0 0},clip]{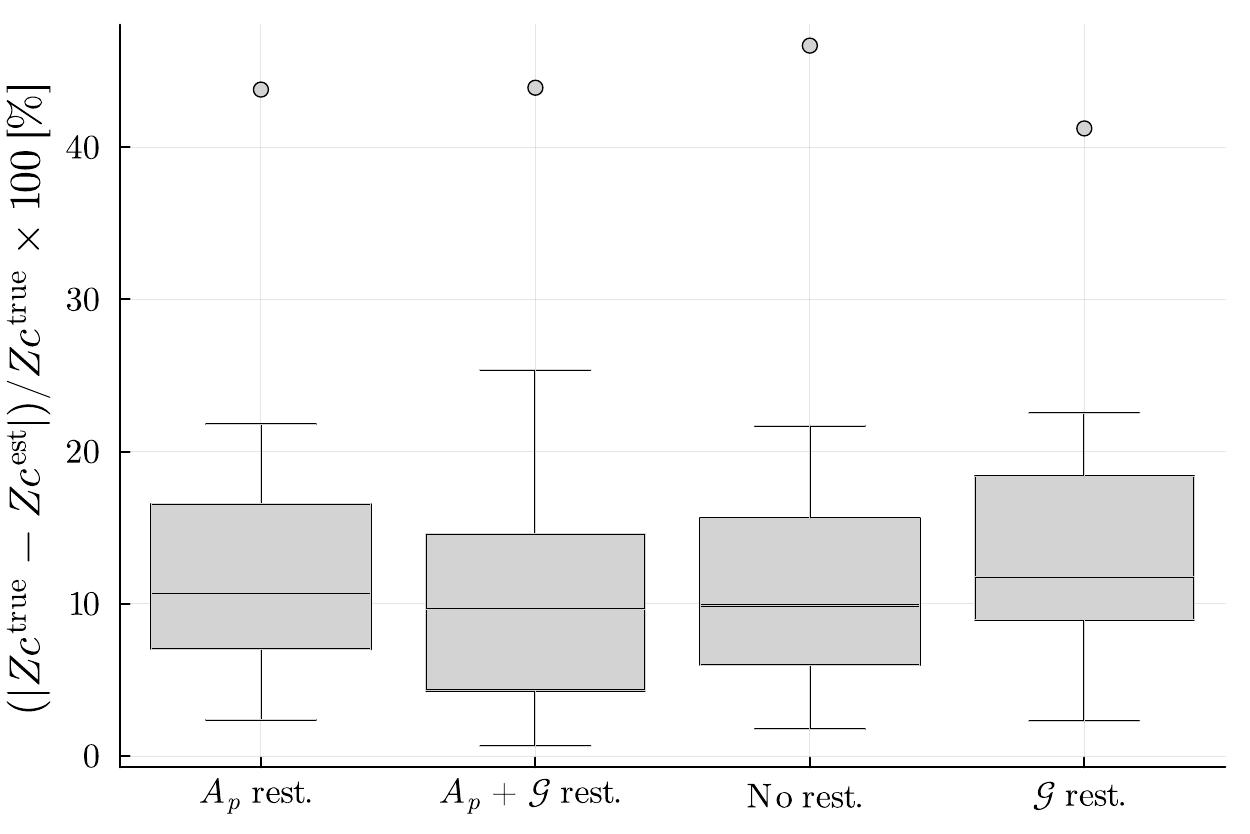}
\\
\includegraphics[width=.31\textwidth]{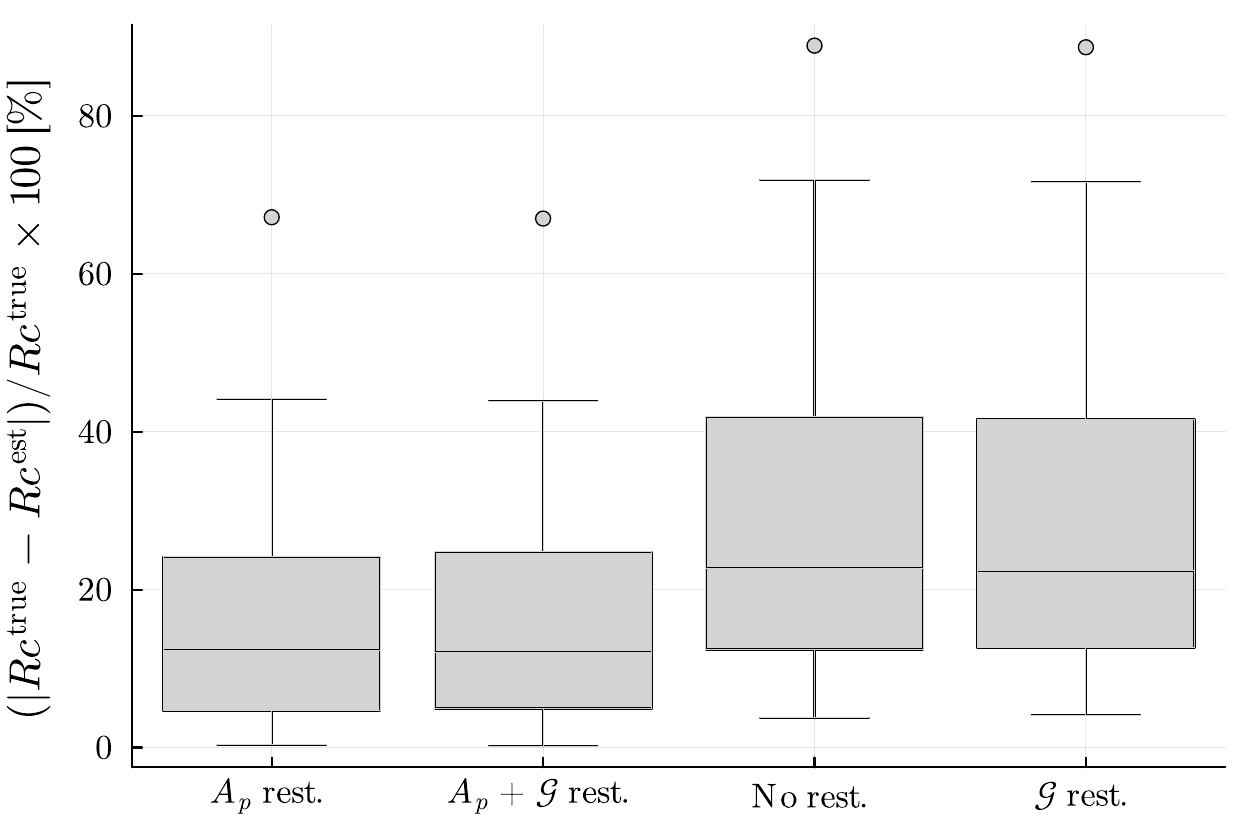}
\includegraphics[width=.31\textwidth]{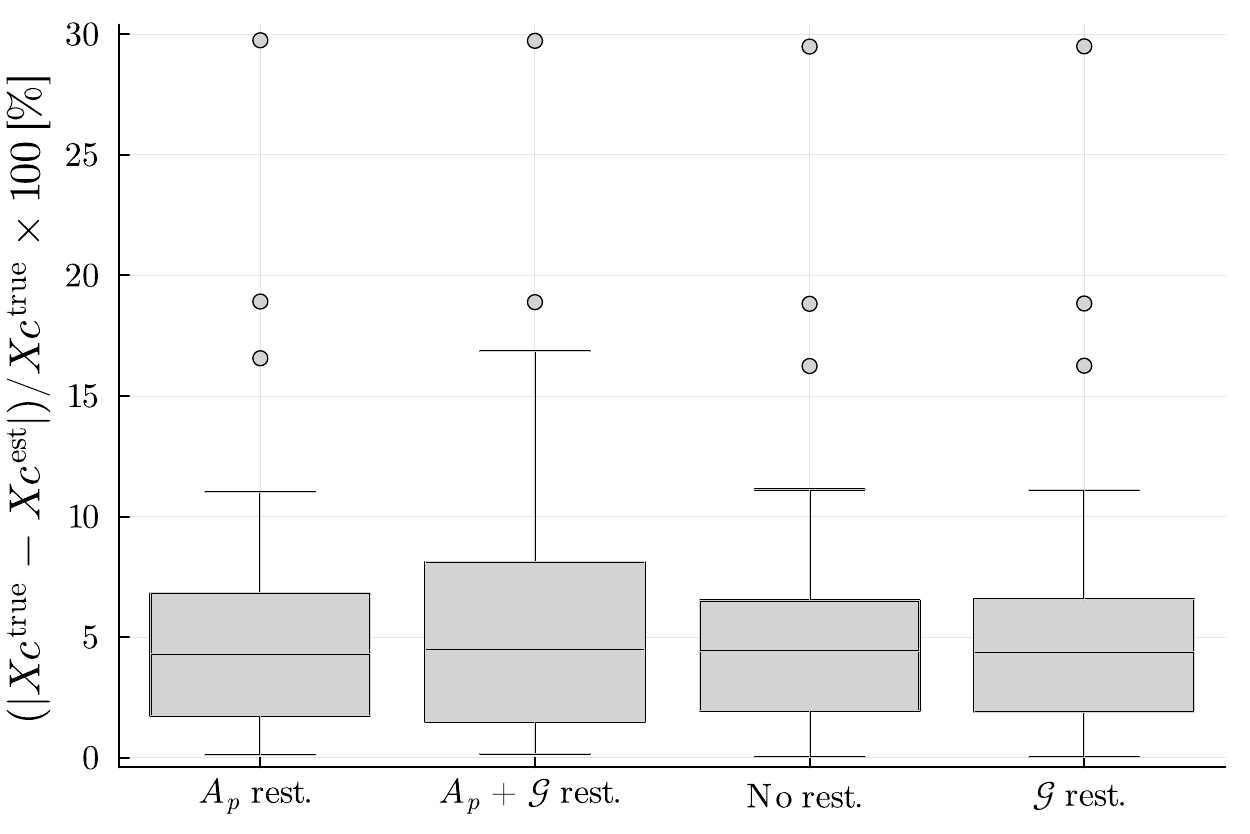}
\includegraphics[width=.31\textwidth]{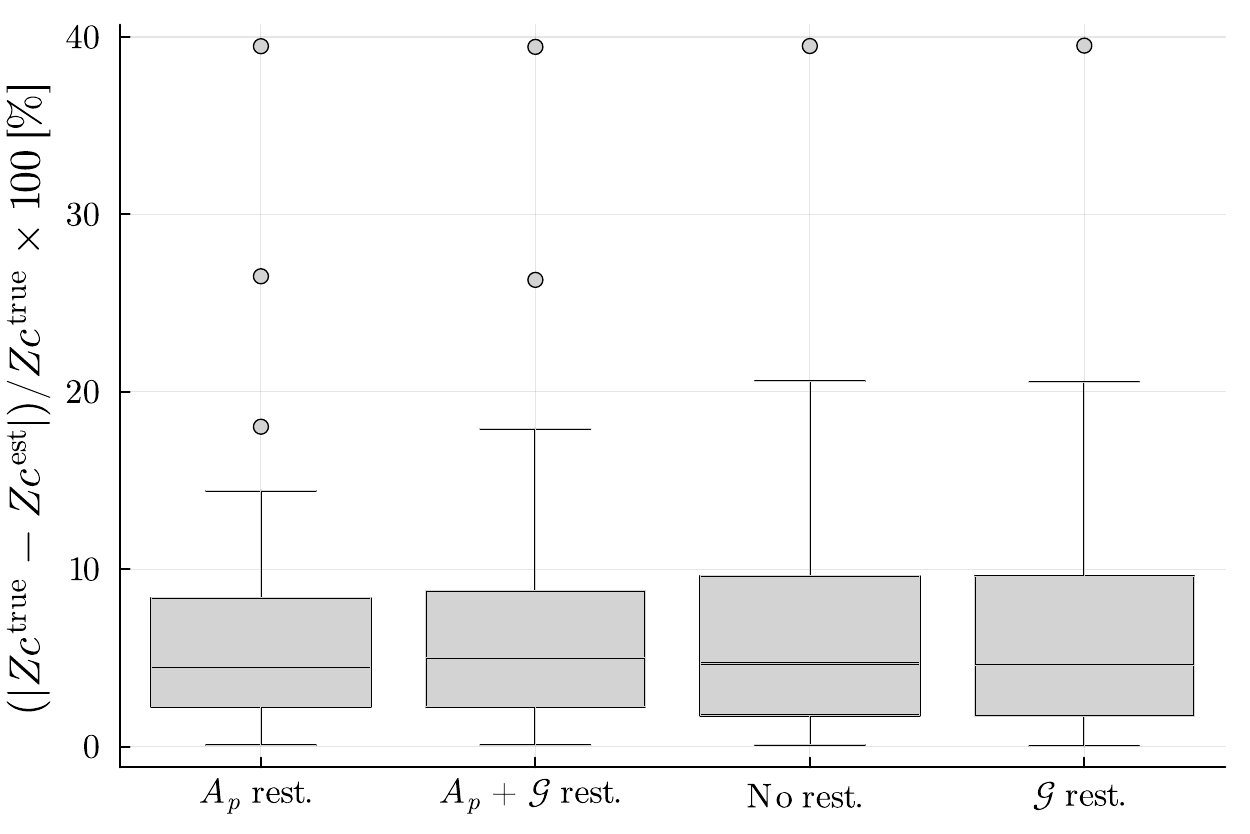}
\caption{Impedance estimation accuracy; the three top figures are for the \textbf{30L case}, the three bottom ones for the \textbf{EU case}. The left-most figures show the percentage difference between the true ($R_c^{true}$) and the estimated ($R_c^{est}$) cumulative resistances. In the middle, the same is shown for the cumulative reactance and to the right for their complex sum (cumulative impedance).}
\label{fig:impedance-30l-ug}
\end{figure*}

\begin{figure*}[t]
\centering
{
\includegraphics[width=.33\textwidth]{result_figures/30l_ug_cumulative_Xc_diff_boxplot_constr_paper_pm1.pdf}
}
{
\includegraphics[width=.33\textwidth]{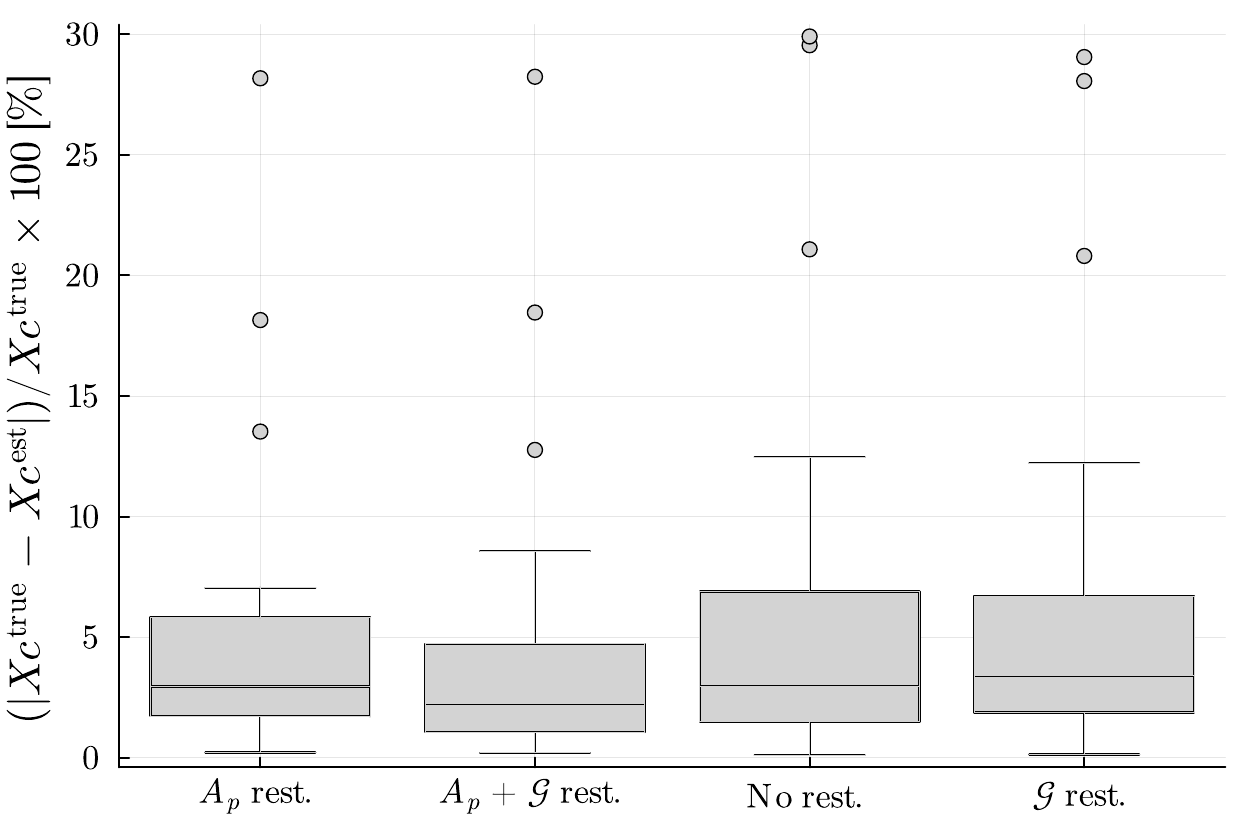}
}
\caption{Improvement in reactance estimation in higher-load conditions (left figure: no load multiplication; right figure: Loads $\times$ 3) - \textbf{30L case}.}
\label{fig:load-multiplier}
\end{figure*}

\subsection{Computational time}
 is reported in Table~\ref{tab:comp-time}. The use of constraints generally reduces solve times, in particular for the slower EU case, proving that the addition of domain knowledge is valuable in this sense. Even when no restrictions are added, the structure imposed by Carson's equations results in computation times that are one order of magnitude shorter than the ``direct" IE in our previous work~\cite{Vanin2023IE}, which took 13603\,s to learn ``smaller" $3 \times 3$ impedance models for the EU case. The addition of further constraints reduces 
the computational time by an additional order of magnitude. The fact that our IE method is less degenerate avoids convergence issues even in sizable networks, which is reassuring in terms of quality of the solution and towards its integration in industrial tools. We note that EU case is larger (it presents more branches, nodes and users) than average (and most) European low voltage feeder, which are usually equipped with 250~kVA transformers design to serve $\approx$40 users in total (i.e., per network, not per feeder). Our IE method can be applied to each feeder separately, without loss of accuracy. Furthermore, for unconventionally large networks that might result in computational challenges, decompositions such as that in~\cite{Vanin2024PI} can be adopted if needed.

\subsection{Cumulative impedance estimation accuracy} is shown in Fig.~\ref{fig:impedance-30l-ug}, in terms of absolute percentage difference between the ground-truth and the IE's estimated values. While estimates of individual branch impedances are known to be inaccurate/meaningless, cumulative impedances (i.e., the impedances of the ``full path" between each user and the transformer), which ultimately determine the quality of the power flow-based calculations, may be accurate~\cite{Vanin2023IE, 9858017, Li2019}.

In the 30L case (Fig.~\ref{fig:impedance-30l-ug}), conductor self-resistances (R) estimates present large outliers, as multi-conductor cables are more reactive than resistive, R is harder to estimate. The higher accuracy in the reactance (X) compensates for those, and results in mean impedance (Z) errors around 10\%. 

The IE accuracy strongly depends on the available data. As mentioned before, increased power flows imply increased voltage drops, and lead to higher signal to noise ratio of voltage measurements. Fig.~\ref{fig:load-multiplier} illustrates how multiplying the power demand by a factor of three considerably improves the impedance estimates. As increased loading is expected as a consequence of electrification, there is clear potential for high-quality IE in the future. Similarly, the IE is more accurate for the EU case, as it has nearly double the amount of users of the 30L case, and hence higher demand.

\begin{table}[t]
\centering
\caption{Time, in seconds, for ipopt to solve to tolerance 1E-8.}\label{tab:comp-time}
\begin{tabular}{r|cc}
    & \multicolumn{1}{c}{30L} & \multicolumn{1}{c}{EU}  \\ \hline
\texttt{No rest.}                & 201.61 & 1761.37    \\
$\texttt{A}_\texttt{p}$~\texttt{rest.}
         & 142.19 &  982.47  \\
\texttt{$\mathcal{G}$ rest.}     & 132.23 &  561.38   \\
\texttt{$\mathcal{G}$+}$\texttt{A}_\texttt{p}$~\texttt{rest.} & 147.85 &  349.15 \\
\hline
\end{tabular}
\end{table}

\subsection{Power flow validation} 
\begin{figure}[t]
\centering
\includegraphics[width=0.4\textwidth]{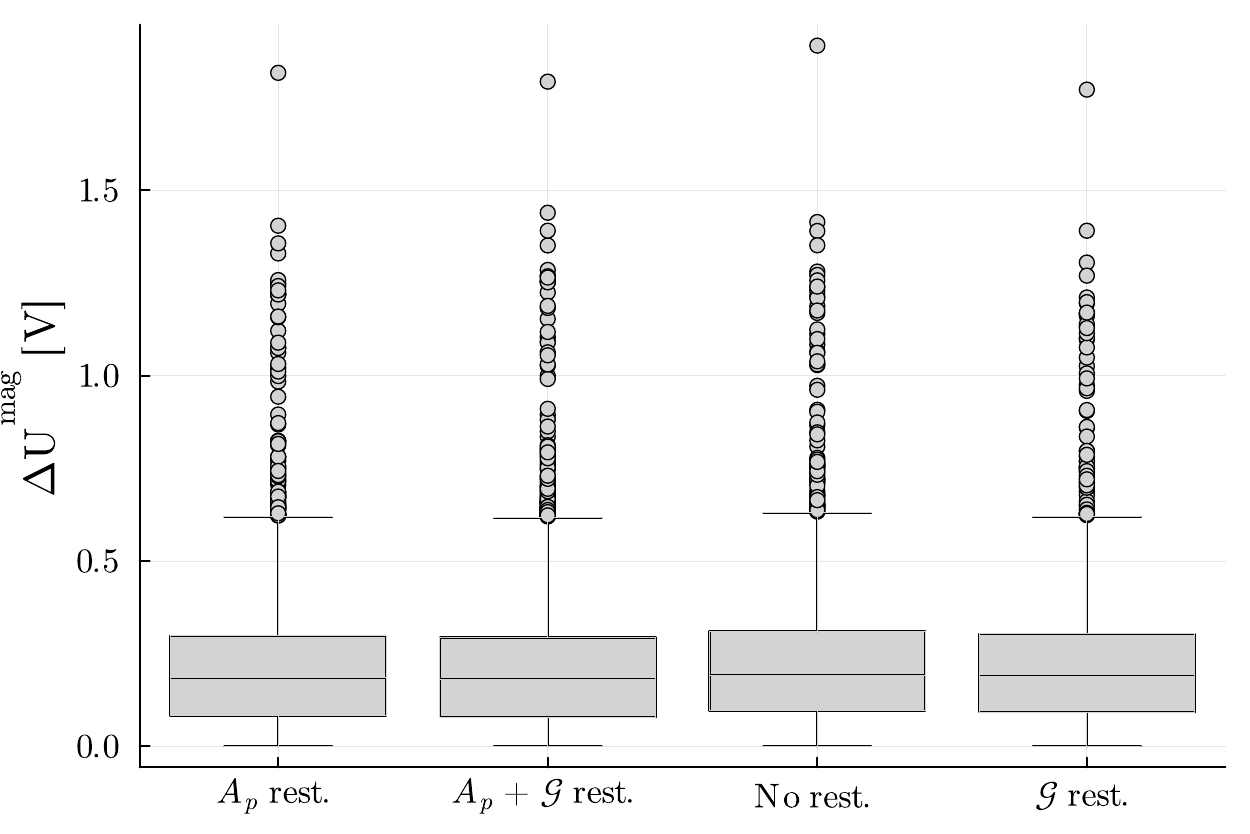}
    \caption{Power flow validation for the \textbf{30L case}.}
    \label{fig:pf-30l-ug}
\end{figure}

\begin{figure}[h!]
\centering
\includegraphics[width=0.4\textwidth]{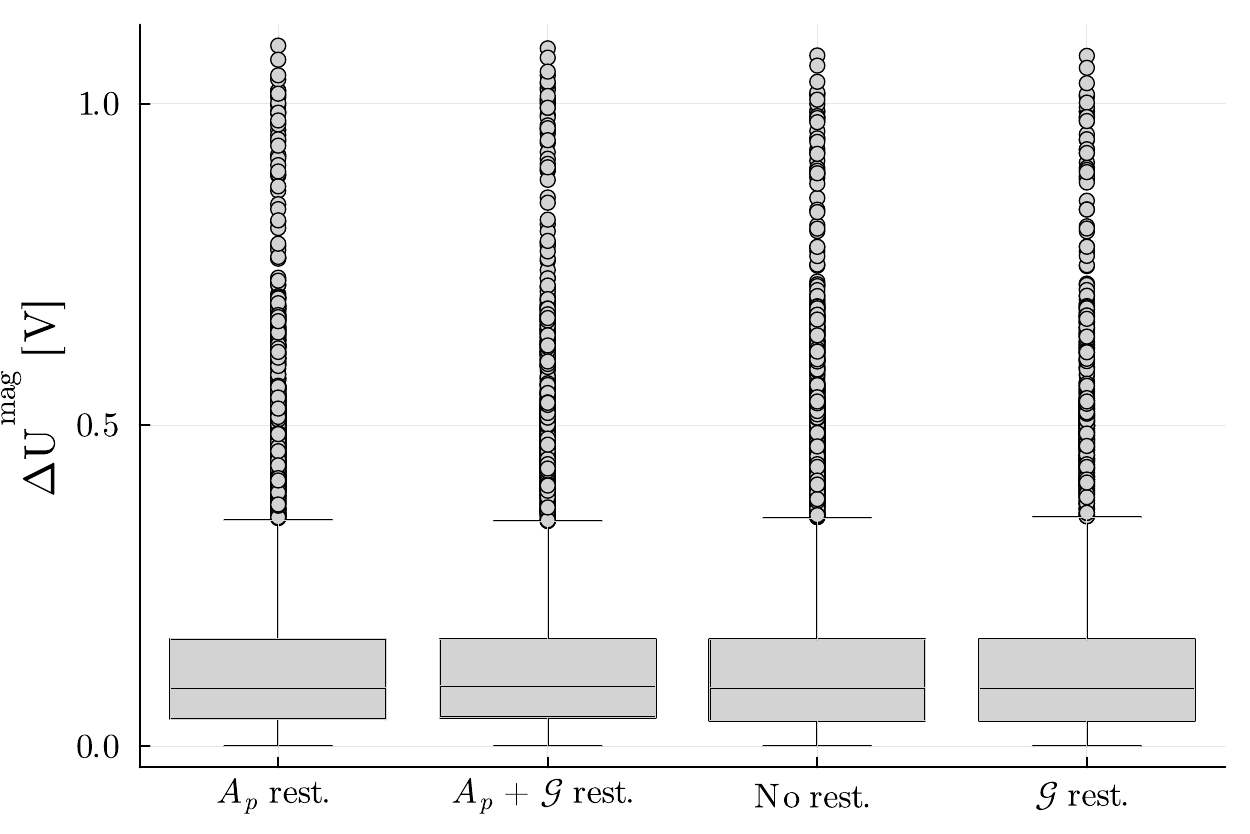}
    \caption{Power flow validation for the \textbf{EU case}.}
    \label{fig:pf-eu-ug}
\end{figure}

analyses the difference in simulation results (on the validation set) using the learned impedances and the ground-truth impedances. Fig.~\ref{fig:pf-30l-ug}-\ref{fig:pf-eu-ug} show (phase-to-ground) absolute voltage magnitude differences $\Delta \text{U}^\text{mag}$ (one dot per bus per time step is shown). The median error is always well below 0.5~V, and the overall accuracy of the learned impedance model appears sufficient. The use of different sets of constraints does not have a noticeable impact on the power flow validation, whereas it did have a moderately positive impact on the cumulative impedance values estimation (i.e., $A_p $ \texttt{rest.} has lower percentage differences than less constrained cases). Same power flow results despite different IE results occur because of the existence of multiple equivalent solutions, as discussed before, and shows how enforcing domain knowledge reduces overfitting on the estimated impedances even though it may not have a visible impact on the power flows. The IE accuracy with Carson is overall superior to our previous work~\cite{Vanin2023IE}.

\begin{figure}[h!]
\centering
\includegraphics[width=0.4\textwidth]{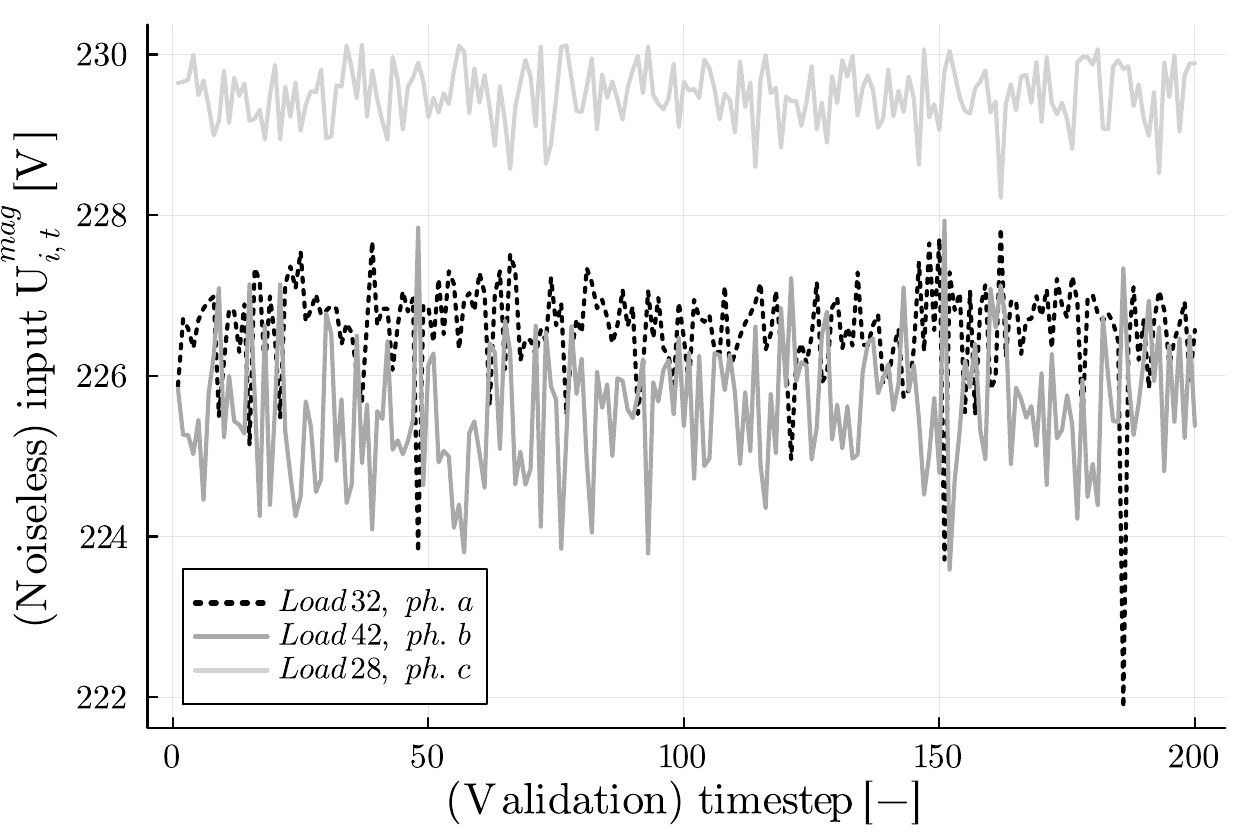}
    \caption{Phase-to-neutral (ground-truth) voltage magnitudes for three loads during the validation timeseries.}
    \label{fig:voltage-timeseries}
\end{figure}

For a reference, Fig.~\ref{fig:voltage-timeseries} shows the timeseries voltage magnitude values for three users (on different phases). These are the $U^{mag}_{i,t}$ from the ground-truth as indicated to the right of Fig.~\ref{fig:flowchart}. Fig.~\ref{fig:voltage-timeseries} shows that the variations are larger than the $\Delta U^{mag}$ in the PF validation figures, i.e., the ``real" input voltage is not flat and the model from IE can capture nontrivial variations.

\begin{figure}[h!]
\centering
\includegraphics[width=.4\textwidth]{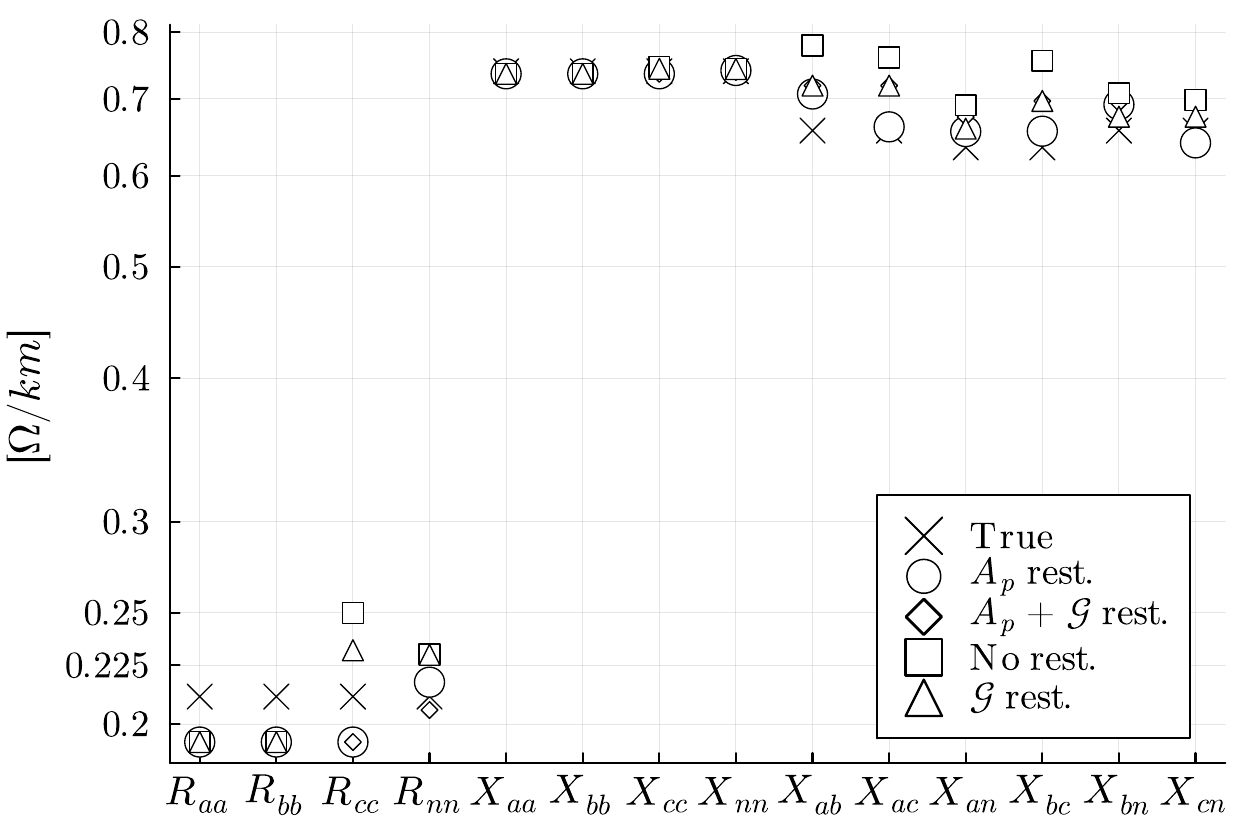}
\includegraphics[width=.4\textwidth]{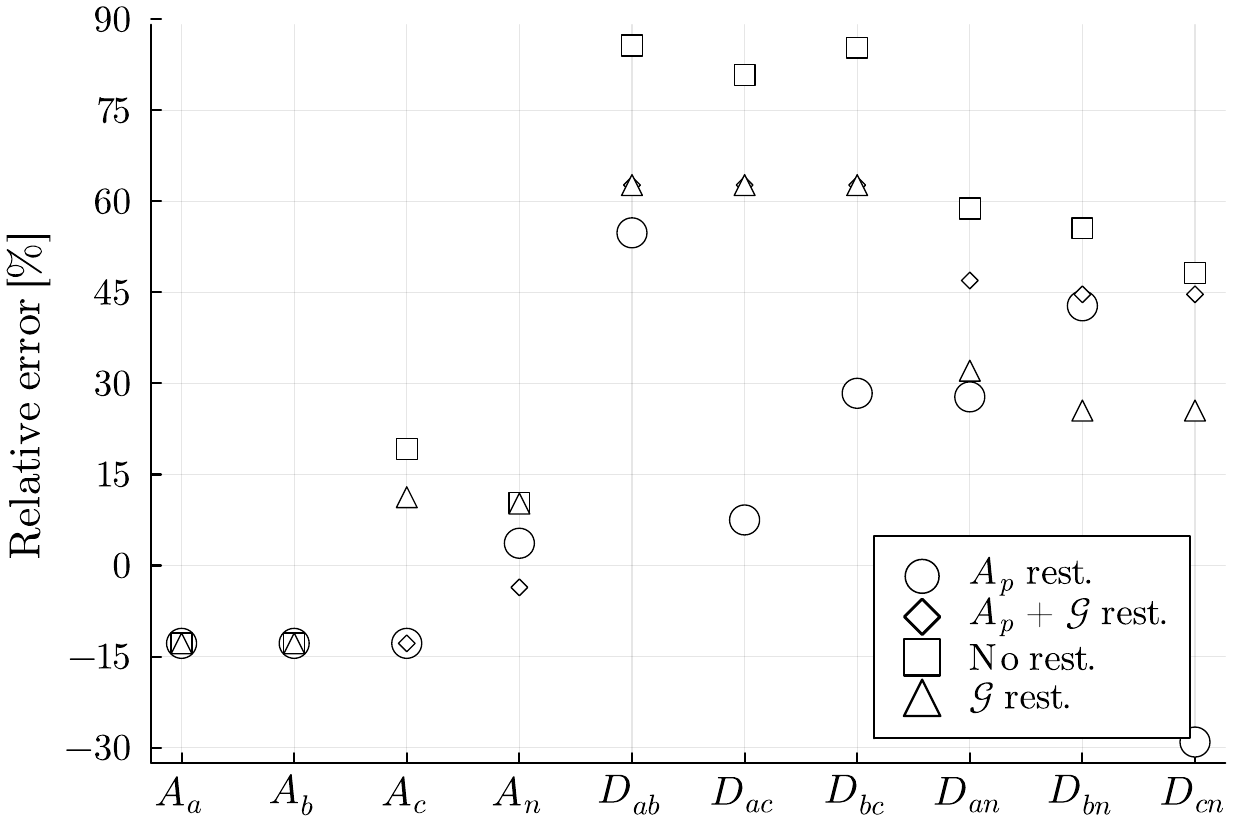}
    \caption{Estimated $\mathbf{Z}^{\text{nom}}$ entries for cable uglv\_240al\_xlpe.}
    \label{fig:uglv_240al_xlpe}
\end{figure}

\begin{figure}[t]
\centering
\includegraphics[width=.4\textwidth]{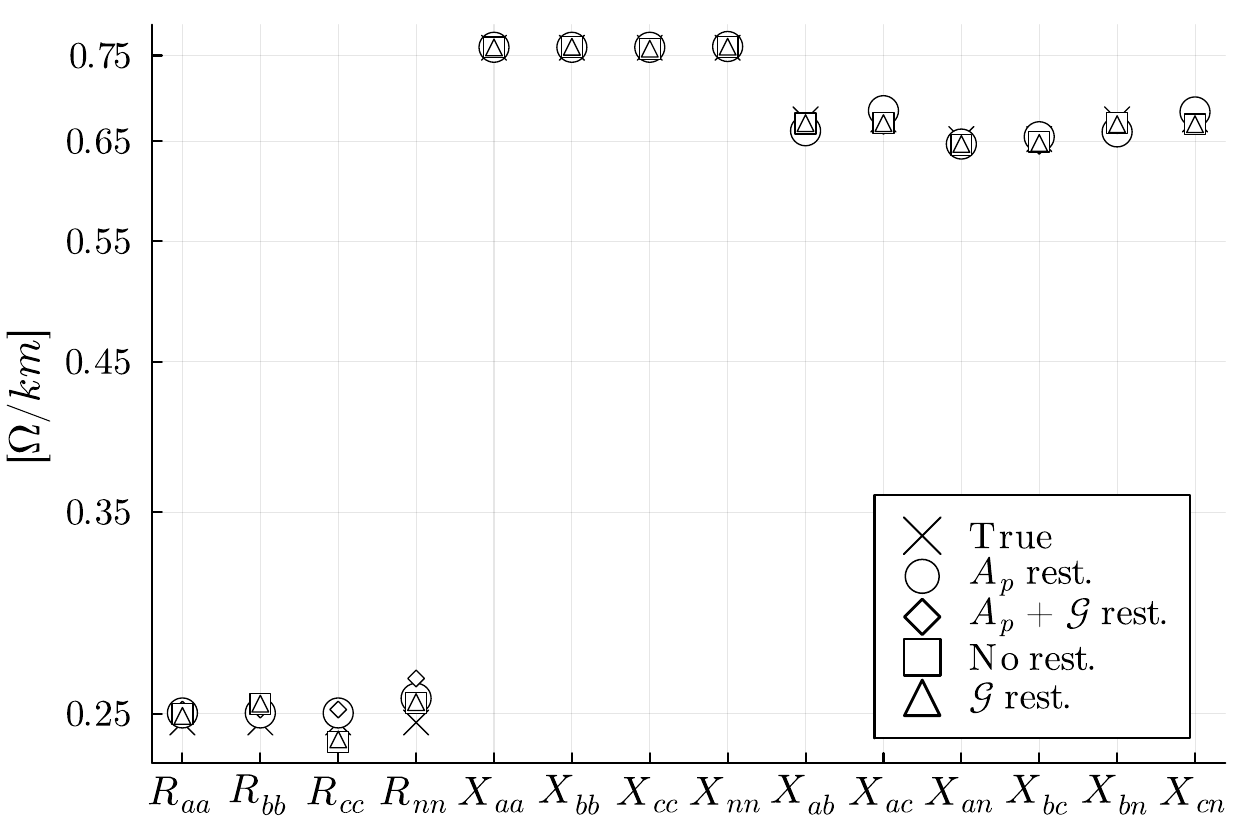}
\includegraphics[width=.4\textwidth]{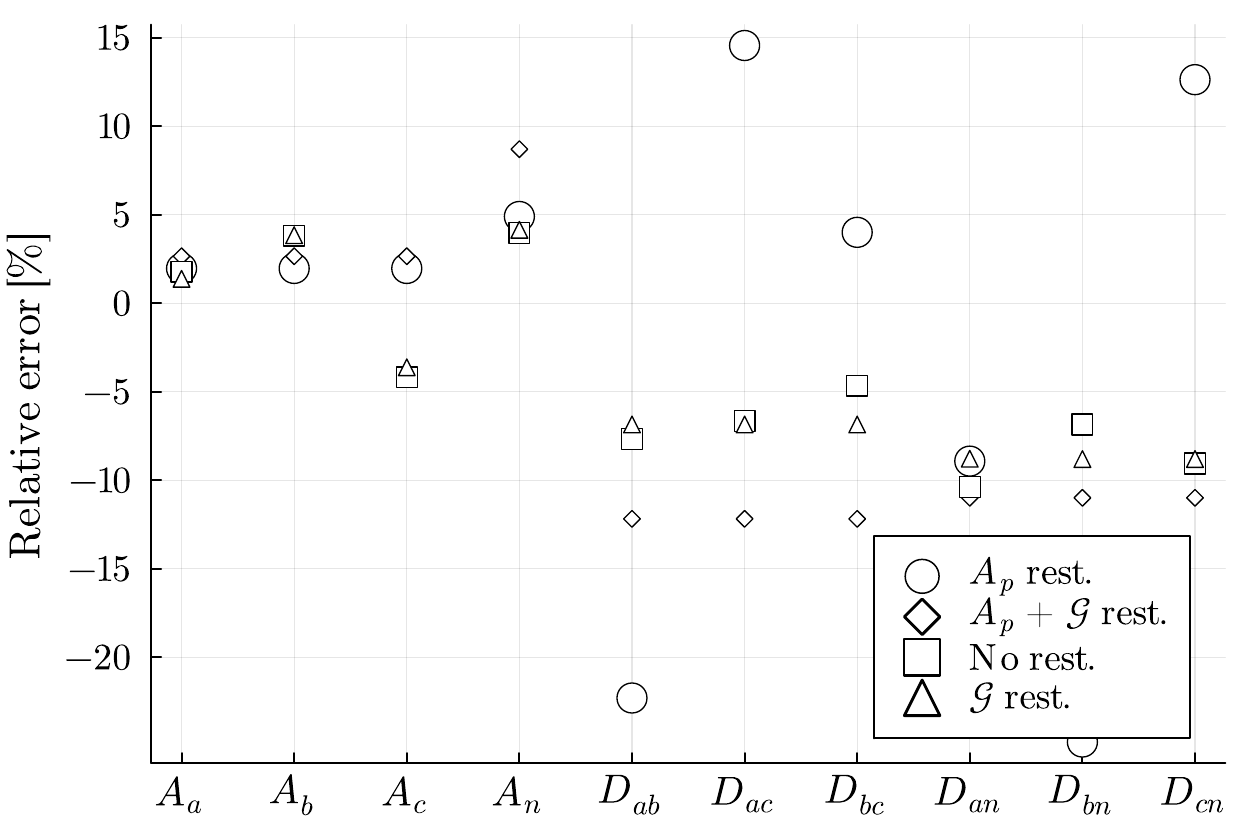}
    \caption{Estimated $\mathbf{Z}^{\text{nom}}$ entries for cable uglv\_120cu\_xlpe.}
    \label{fig:uglv_120cu_xlpe}
\end{figure}

\begin{figure}[t]
\centering

\includegraphics[width=.4\textwidth]{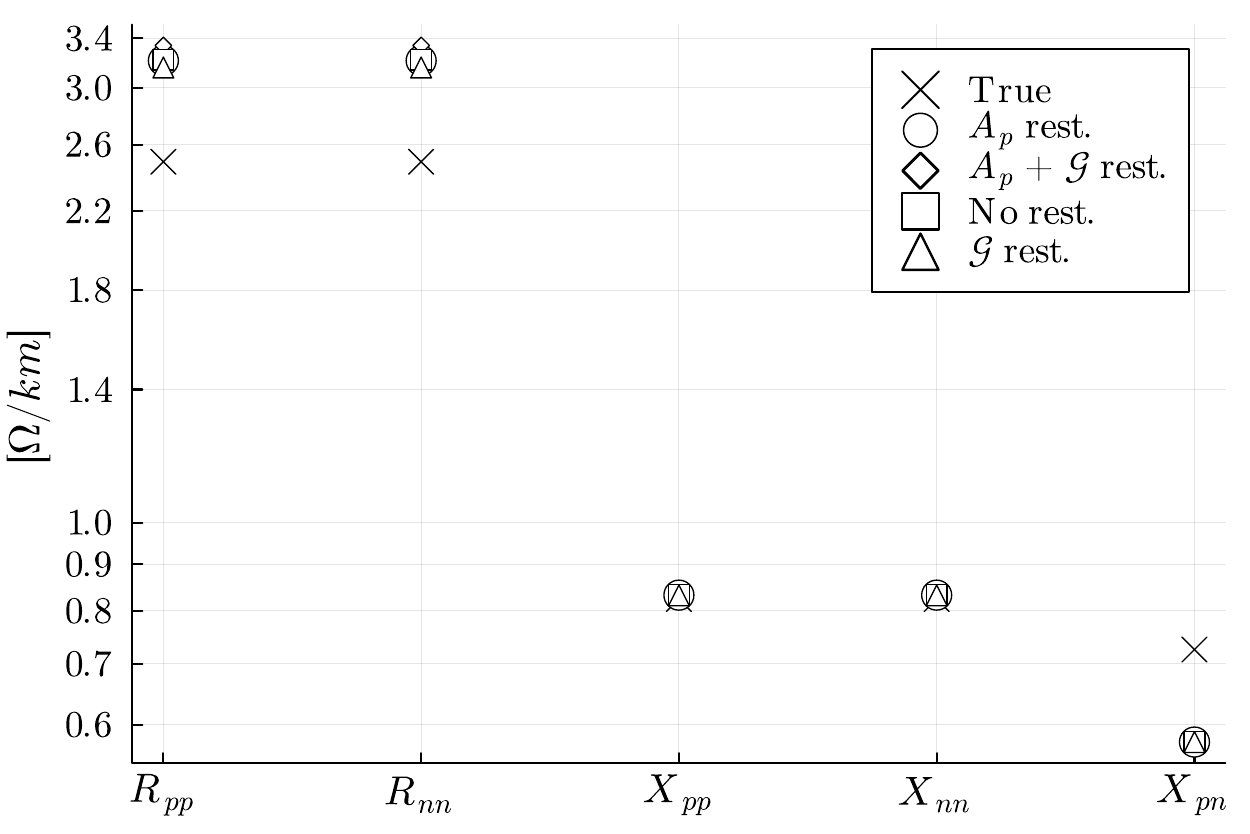}
\includegraphics[width=.4\textwidth]{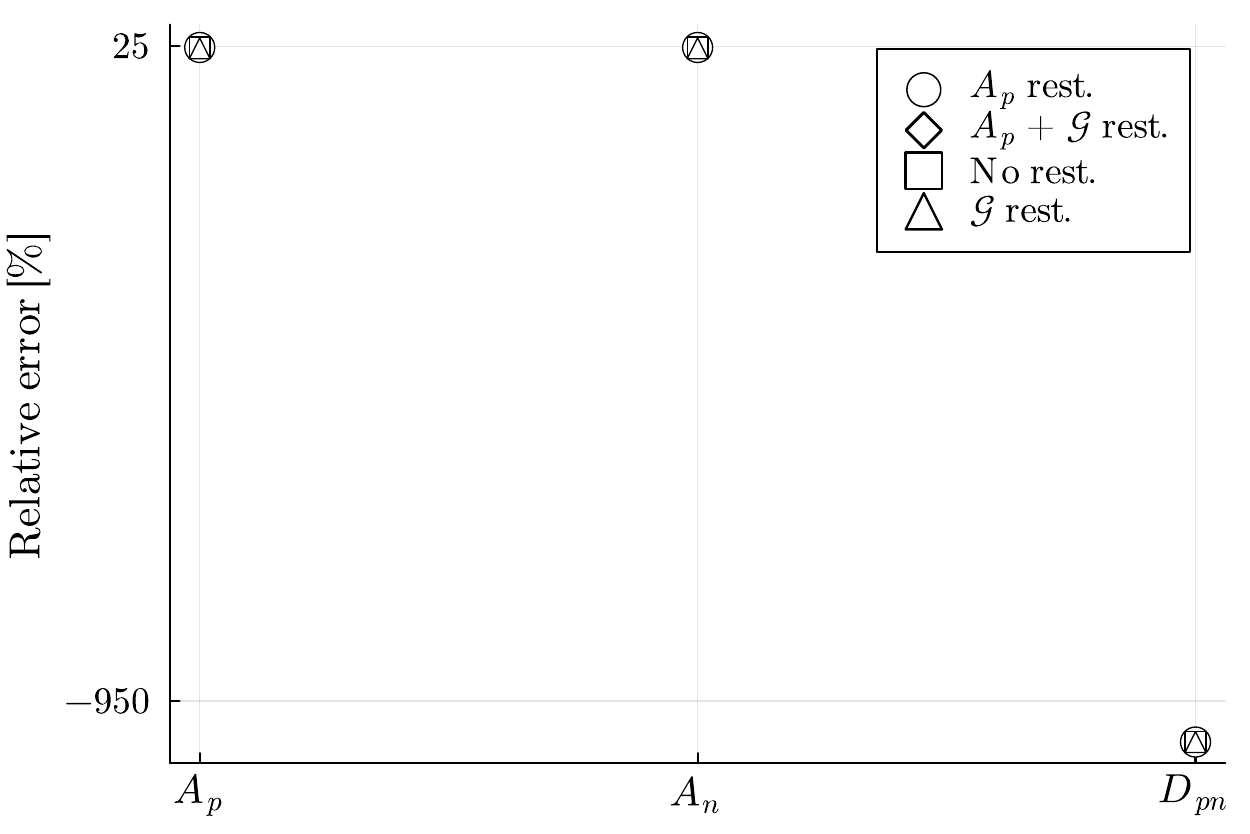}
    \caption{Estimated $\mathbf{Z}^{\text{nom}}$ entries for cable ugsc\_16al\_xlpe.}
    \label{fig:ugsc_16al_xlpe}
\end{figure}

\subsection{Construction code estimation accuracy} Fig.~\ref{fig:uglv_240al_xlpe},~\ref{fig:uglv_120cu_xlpe} show the construction code impedance entries estimated for the two multi-conductor cables, as well as their corresponding cross-section and conductor distances. The entry estimates are accurate, but the geometrical properties often present a high error with respect to their true value. This is again due to the existence of multiple combinations of geometrical properties resulting in the same solution. The results for \textsc{uglv\_120cu\_xlpe} are better than \textsc{uglv\_240al\_xlpe} case because the former is located in the higher-load EU feeder.

While geometry estimates might not be realistic, and as such hard to ``match" to a specific construction code, the $\mathbf{Z}^{\text{nom}}$ entries could be reduced to positive and zero sequence values, and used as input to the inverse Carson problem in~\cite{Tam2024}. 

Service cables experience considerably lower power flows, so while the same conclusions apply, their relative results are usually worse, as illustrated in Fig.~\ref{fig:ugsc_16al_xlpe}. The other service cables present similar figures, omitted for space reasons.

Finally, we observe that individual branch lengths cannot be accurately estimated, which is again due to the presence of multiple solutions, and was expected: even in line length estimation~\cite{Vanin2023IE}, in which nominal impedances are known and only branch lengths are variables (see Table~\ref{tab:variables-fig}), correct individual lengths cannot be recovered. Future work will explore iterative variable assignment processes to try and overcome this effect.

\section{Conclusion and Future Work}\label{s:conclusion}
This paper proposes a state-of-the-art method exploiting Carson's equations to perform joint state and impedance matrix estimation in unbalanced distribution networks with up to four wires, using time series data with realistic features, similar to those of residential smart meters. 

Impedance estimation improves the quality of the network data that utilities use to perform physics-based power systems calculations, like (optimal) power flow. Such calculations are crucial to adequately plan and operate modern distribution networks with high levels of electrification and distributed generation. As utilities usually only have rough estimates of impedance matrices, improving their values increases the reliability of the calculation results. However, as discussed in our paper, the identification of exact impedance values appears impossible given the features of present low voltage networks and their telemetry. Thus, developing mathematical models that account for parameter uncertainty in distribution system calculations is recommended.

Our impedance estimation approach is more generic than existing methods, and up to two orders of magnitude faster than our previous work, 
thanks to Carson's equations and the dimensional reduction they enable. The data and code have been made openly available. 

In addition to computational advantages, Carson's equations allow a better exploitation of domain knowledge, which in turn reduces the risk of overfitting, improves impedance estimates and potentially infers geometrical/material properties that are more informative and easy to verify than equivalent impedance models alone. Retrieving these properties may allow extra applications that are impossible through equivalent impedances, e.g., recalculating impedance values for other frequencies to perform harmonic studies.

\section*{Acknowledgment}
This work received funding from the Agency for Innovation and Entrepreneurship of the Flemish Government (VLAIO) and Flux50, through the strategic research project IMPRO-CAP (Grant N° HBC 2022 0733) and is supported by the Australian Department of Climate Change, Energy, the Environment and Water under the International Clean Innovation Researcher Networks (ICIRN) program grant number ICIRN000072. The collaboration involved a research visit that received a grant from the Science Foundation: Flanders (FWO), Grant V420224N.

\bibliographystyle{IEEEtran}
\bibliography{IEEEabrv,Bibliography_abb}

\begin{thebibliography}{10}
\providecommand{\url}[1]{#1}
\csname url@samestyle\endcsname
\providecommand{\newblock}{\relax}
\providecommand{\bibinfo}[2]{#2}
\providecommand{\BIBentrySTDinterwordspacing}{\spaceskip=0pt\relax}
\providecommand{\BIBentryALTinterwordstretchfactor}{4}
\providecommand{\BIBentryALTinterwordspacing}{\spaceskip=\fontdimen2\font plus
\BIBentryALTinterwordstretchfactor\fontdimen3\font minus \fontdimen4\font\relax}
\providecommand{\BIBforeignlanguage}[2]{{%
\expandafter\ifx\csname l@#1\endcsname\relax
\typeout{** WARNING: IEEEtran.bst: No hyphenation pattern has been}%
\typeout{** loaded for the language `#1'. Using the pattern for}%
\typeout{** the default language instead.}%
\else
\language=\csname l@#1\endcsname
\fi
#2}}
\providecommand{\BIBdecl}{\relax}
\BIBdecl

\bibitem{Carson1926}
J.~R. Carson, ``Wave propagation in overhead wires with ground return,'' \emph{The Bell System Technical Journal}, vol.~5, no.~4, pp. 539--554, 1926.

\bibitem{kersting2018distribution}
W.~H. Kersting, ``Distribution system modeling and analysis.''\hskip 1em plus 0.5em minus 0.4em\relax CRC press, 2018.

\bibitem{GethCIRED2023}
F.~Geth, M.~Vanin, and D.~Van~Hertem, ``Data quality challenges in existing distribution network datasets,'' in \emph{CIRED 2023}, Rome, Italy.

\bibitem{nyserda}
``Fundamental research challenges for distribution state estimation to enable high-performing grids,'' NYSERDA, Technical Report, 2018.

\bibitem{Cherot2023}
G.~{Chérot, et al.}, ``Misestimation of impedance values within a distribution network optimal power flow,'' in \emph{IEEE PowerTech}, 2023, pp. 1--6.

\bibitem{Cavraro2019CNS}
G.~Cavraro and V.~Kekatos, ``Inverter probing for power distribution network topology processing,'' \emph{IEEE Control Netw. Syst.}, vol.~6, no.~3, pp. 980--992, 2019.

\bibitem{Vanin2023IE}
M.~Vanin, F.~Geth, R.~D’hulst, and D.~{Van Hertem}, ``Combined unbalanced distribution system state and line impedance matrix estimation,'' \emph{Int. J. Electr. Power Energy Syst.}, vol. 151, p. 109155, 2023.

\bibitem{Zhang2021}
J.~{Zhang, et al.}, ``Distribution network admittance matrix estimation with linear regression,'' \emph{IEEE Trans. Power Syst.}, vol.~36, no.~5, pp. 4896--4899, 2021.

\bibitem{Zhang2020Topology}
J.~Zhang, Y.~Wang, Y.~Weng, and N.~Zhang, ``Topology identification and line parameter estimation for non-pmu distribution network: A numerical method,'' \emph{IEEE Trans. Smart Grid}, vol.~11, no.~5, pp. 4440--4453, 2020.

\bibitem{Guo2022}
Y.~{Guo, et al.}, ``Distribution grid modeling using smart meter data,'' \emph{IEEE Trans. Power Syst.}, vol.~37, no.~3, pp. 1995--2004, 2022.

\bibitem{Marulli2021}
D.~{Marulli, et al.}, ``Reconstruction of low-voltage networks with limited observability,'' in \emph{IEEE PES ISGT Europe}, 2021, pp. 1--5.

\bibitem{Kapoor2024}
S.~{Kapoor, et al.}, ``Maximum likelihood estimation of state variables and line parameters in distribution grid with a non-linear model,'' \emph{Authorea Preprints}, 2024.

\bibitem{Sang2024pscc}
P.~Sang and A.~Pandey, ``Circuit-theoretic joint parameter-state estimation—balancing optimality and ac feasibility,'' \emph{Electr. Power Syst. Res.}, vol. 235, p. 110637, 2024.

\bibitem{Mittal2024}
S.~Mittal, P.~Pareek, and A.~Verma, ``Distribution line parameters estimation framework with correlated injections using smart meter measurements,'' \emph{Electr. Power Syst. Res.}, vol. 228, p. 110083, 2024.

\bibitem{Li2022}
X.~Li, S.~Wang, and Z.~Lu, ``Reverse identification method of line parameters in distribution network with multi-t nodes based on partial measurement data,'' \emph{Electr. Power Syst. Res.}, vol. 204, p. 107691, 2022.

\bibitem{GuptaPaolone}
R.~{Gupta, et al.}, ``Compound admittance matrix estimation of three-phase untransposed power distribution grids using synchrophasor measurements,'' \emph{IEEE Trans. Instrum. Meas.}, vol.~70, pp. 1--13, 2021.

\bibitem{Moffat2020}
K.~Moffat, M.~Bariya, and A.~Von~Meier, ``Unsupervised impedance and topology estimation of distribution networks—limitations and tools,'' \emph{IEEE Trans. Smart Grid}, vol.~11, no.~1, pp. 846--856, 2020.

\bibitem{ClaeysCIRED2021}
S.~{Claeys, et al.}, ``Line parameter estimation in multi-phase distribution networks without voltage angle measurements,'' in \emph{CIRED 2021}, pp. 1186--1190.

\bibitem{Dutta}
R.~{Dutta, et al.}, ``Parameter estimation of distribution lines using scada measurements,'' \emph{IEEE Trans. Instrum. Meas.}, vol.~70, pp. 1--11, 2021.

\bibitem{WangPMAPS2020}
W.~Wang and N.~Yu, ``Parameter estimation in three-phase power distribution networks using smart meter data,'' in \emph{Int. Conf. Probabilistic Methods Appl. Power Syst.}, 2020, pp. 1--6.

\bibitem{Jaepil2022}
J.~{Ban, et al.}, ``{AMI} data-driven strategy for hierarchical estimation of distribution line impedances,'' \emph{IEEE Trans. Power Del.}, pp. 1--13, 2022.

\bibitem{Peppanen2016}
J.~{Peppanen, et al.}, ``Distribution system model calibration with big data from {AMI} and {PV} inverters,'' \emph{IEEE Trans. Smart Grid}, vol.~7, no.~5, pp. 2497--2506, 2016.

\bibitem{Yang2022}
N.-C. {Yang, et al.}, ``Three-phase feeder parameter estimation using radial basis function neural networks and multi-run optimisation method with bad data preparation,'' \emph{IET GTD}, vol.~16, no.~2, pp. 351--363, 2022.

\bibitem{Costa2022}
L.~{Costa, et al.}, ``Identification and correction of transmission line parameter errors using scada and synchrophasor measurements,'' \emph{Int. J. Electr. Power Energy Syst.}, vol. 135, p. 107509, 2022.

\bibitem{Short2013}
T.~A. Short, ``Advanced metering for phase identification, transformer identification, and secondary modeling,'' \emph{IEEE Trans. Smart Grid}, vol.~4, no.~2, pp. 651--658, 2013.

\bibitem{Lave2019}
M.~{Lave, et al.}, ``Distribution system parameter and topology estimation applied to resolve low-voltage circuits on three real distribution feeders,'' \emph{IEEE Trans. Sustain. Energy}, vol.~10, no.~3, pp. 1585--1592, 2019.

\bibitem{Lin2018framework}
Y.~Lin and A.~Abur, ``A new framework for detection and identification of network parameter errors,'' \emph{IEEE Trans. Smart Grid}, vol.~9, no.~3, pp. 1698--1706, 2018.

\bibitem{bible}
A.~Abur and A.~G\'{o}mez-Exp\'{o}sito, \emph{Power System State Estimation: Theory and Implementation}, Book, CRC Press, 2004.

\bibitem{Alam2024}
M.~T. Alam and B.~Das, ``Decoupled state and line parameter estimation in three-phase unbalanced distribution system,'' \emph{Electr. Power Syst. Res.}, vol. 229, p. 110127, 2024.

\bibitem{9858017}
Y.~Yuan, S.~H. Low, O.~Ardakanian, and C.~J. Tomlin, ``Inverse power flow problem,'' \emph{IEEE Trans. Control Network Syst.}, pp. 1--12, 2022.

\bibitem{Li2019}
T.~Li, L.~Werner, and S.~H. Low, ``Learning graph parameters from linear measurements: Fundamental trade-offs and application to electric grids,'' in \emph{IEEE 58th Conf. Decision Control}, 2019, pp. 6554--6559.

\bibitem{Brouillon2024}
J.-S. Brouillon, K.~Moffat, F.~Dörfler, and G.~Ferrari-trecate, ``Power grid parameter estimation without phase measurements: Theory and empirical validation,'' \emph{Power Systems Comp. Conf.}, 2024.

\bibitem{Cunha2020}
V.~{Cunha, et al.}, ``Automated determination of topology and line parameters in low voltage systems using smart meters measurements,'' \emph{IEEE Trans. Smart Grid}, vol.~11, no.~6, pp. 5028--5038, 2020.

\bibitem{cleenwerck2022}
R.~{Cleenwerck, et al.}, ``An approach to the impedance modelling of low-voltage cables in digital twins,'' \emph{Electric Power Syst. Res.}, vol. 210, p. 108075, 2022.

\bibitem{Kersting2011}
W.~H. Kersting and R.~K. Green, ``The application of carson's equation to the steady-state analysis of distribution feeders,'' in \emph{IEEE/PES Power Syst. Conf. Expo.}, 2011, pp. 1--6.

\bibitem{Tam2024}
C.~H. {Tam, et al.}, ``The inverse carson's equations problem: Definition, implementation and experiments,'' \emph{preprint: arXiv:2404.08210}, 2024.

\bibitem{CLAEYS2022108522}
S.~Claeys, F.~Geth, and G.~Deconinck, ``Optimal power flow in four-wire distribution networks: Formulation and benchmarking,'' \emph{Electr. Power Syst. Res.}, vol. 213, p. 108522, 2022.

\bibitem{Lacroix1995}
\BIBentryALTinterwordspacing
B.~Lacroix and C.~R., ``Earthing systems worldwide and evolutions,'' no. 173, 1995. [Online]. Available: \url{https://www.studiecd.dk/cahiers_techniques/System_earthings_worldwide_and_evolutions.pdf}
\BIBentrySTDinterwordspacing

\bibitem{Rayati2024pscc}
M.~Rayati, M.~Bozorg, and O.~Alizadeh-Mousavi, ``Admittance matrix estimation of radial distribution grids using power and voltage magnitude measurements,'' \emph{accepted for Power Syst. Comp. Conf.,}, 2024.

\bibitem{Urquhart2015}
A.~J. Urquhart and M.~Thomson, ``Series impedance of distribution cables with sector-shaped conductors,'' \emph{IET Gener. Transm. Distr.}, vol.~9, no.~16, pp. 2679--2685, 2015.

\bibitem{Ghaderi2023}
A.~Ghaderi~Baayeh and M.~Kleemann, ``Estimating grounding resistance of medium voltage cables using measured sheaths current,'' in \emph{Int. Conf. Smart Grid Metrology}, 2023, pp. 1--5.

\bibitem{Low2024}
S.~H. Low, ``Reverse kron reduction of multi-phase radial network,'' \emph{preprint: arXiv:2403.17391}, 2024.

\bibitem{geth2022pitfalls}
F.~Geth, ``Pitfalls of zero voltage values in optimal power flow problems,'' \emph{IEEE PES General Meeting}, 2023.

\bibitem{DekaTutorial}
D.~{Deka, et al.}, ``Learning distribution grid topologies: A tutorial,'' \emph{IEEE Trans. Smart Grid}, vol.~15, no.~1, pp. 999--1013, 2024.

\bibitem{Vanin2024PI}
M.~Vanin, T.~Van~Acker, R.~D'hulst, and D.~Van~Hertem, ``Phase identification of distribution system users through a {MILP} extension of state estimation,'' \emph{Elect. Power Syst. Res.}, vol. 229, p. 110107, 2024.

\bibitem{Vanin2022}
M.~Vanin, T.~Van~Acker, R.~D’hulst, and D.~Van~Hertem, ``A framework for constrained static state estimation in unbalanced distribution networks,'' \emph{IEEE Trans. Power Syst.}, vol.~37, no.~3, pp. 2075--2085, 2022.

\bibitem{Rigoni2016}
V.~{Rigoni, et al.}, ``Representative residential lv feeders: A case study for the north west of england,'' \emph{IEEE Trans. Power Syst.}, vol.~31, no.~1, pp. 348--360, 2016.

\bibitem{oedi_4520}
\BIBentryALTinterwordspacing
E.~{Wilson, et al.}, ``End-use load profiles for the u.s. building stock,'' 10 2021. [Online]. Available: \url{https://data.openei.org/submissions/4520}
\BIBentrySTDinterwordspacing

\bibitem{JuMP}
I.~{Dunning, et al.}, ``Ju{MP}: A modeling language for mathematical optimization,'' \emph{SIAM Rev.}, vol.~59, no.~2, pp. 295--320, 2017.

\bibitem{PMD_PSCC}
D.~M. Fobes, S.~Claeys, F.~Geth, and C.~Coffrin, ``{PowerModelsDistribution.jl}: An open-source framework for exploring distribution power flow formulations,'' \emph{Electr. Power Syst. Res.}, vol. 189, p. 106664, 2020.

\bibitem{ipopt}
A.~{W\"{a}chter} and L.~{Biegler}, ``On the implementation of an interior-point filter line-search algorithm for large-scale nonlinear programming,'' \emph{Math. Program.}, vol. 106, no.~1, pp. 25--57, 2006.

\bibitem{HSL}
``{HSL}. {A} collection of fortran codes for large scale scientific computation,'' {A}ccessed: Apr. 2024. [Online]. Available: \url{http://www.hsl.rl.ac.uk/}.

\end{thebibliography}

\end{document}